\documentstyle[12pt]{article}
\input epsf
\makeatletter

\@addtoreset{equation}{section}
\makeatother
\newcommand{\tselea}[1]{\label{#1}}
\newcommand{\tseleq}[1]{\label{#1}}

\newcommand{\tseref}[1]{\ref{#1}} 
\newcommand{\tsecite}[1]{\ \cite{#1}}

\newcommand{\tsebibitem}[1]{\bibitem{#1}} 

\newcommand{\vect}[1]{\cal{\bf #1}}

\newcommand{\cali}[1]{\cal #1}

\font\sans=cmss12
\font\upright=cmu10 scaled\magstep1
\newcommand{\ssf}{\sans}

\newcommand{\M}{\hbox{\upright\rlap{I}\kern 1.7pt M}}
\newcommand{\D}{\hbox{\upright\rlap{I}\kern 1.7pt D}}
\newcommand{\R}{\hbox{\upright\rlap{I}\kern 1.7pt R}}
\newcommand{\Z}{\hbox{\upright\rlap{\ssf Z}\kern 2.7pt {\ssf Z}}}
\newcommand{\SS}{\hbox{\rlap{$\Sigma$}\kern 1pt $\Sigma$}}

\begin{document}

\typeout{--- Title page start ---}

\renewcommand{\thefootnote}{\fnsymbol{footnote}}


\begin{center}{\Large\bf Hydrostatic  pressure of the $O(N)$
$\phi^{4}$ theory in the large $N $ limit} \vskip 1.2cm {\large
Petr Jizba\footnote{E-mail: {\tt petr$@$cm.ph.tsukuba.ac.jp}\\
{\tt $\mbox{\hspace{17mm}}$ p.jizba$@$damtp.cam.ac.uk}}}\vskip 3mm
{\em DAMTP, University of Cambridge, Silver Street, Cambridge, CB3
9EW,
UK\\
and \\
Institute for Theoretical Physics, University of Tsukuba, Ibaraki
305--8571, Japan\footnote{Present address.} }
\end{center}

\addtocounter{footnote}{2}


\vspace{3mm}



\begin{abstract}
\noindent  With non--equilibrium applications in mind we present
in this paper (first in a series of three) a self--contained
calculation of the hydrostatic pressure of the $O(N)~\lambda
\phi^{4}$ theory at finite temperature. By combining the
Keldysh--Schwinger closed--time path formalism with thermal
Dyson--Schwinger equations we compute in the large $N$ limit the
hydrostatic pressure in a fully resumed form. We also calculate
the high--temperature expansion for the pressure (in $D=4$) using
the Mellin transform technique. The result obtained extends the
results found by Drummond {\em et al.}\tsecite{ID1} and
Amelino--Camelia and Pi\tsecite{ACP}. The latter are reproduced in
the limits $m_r(0)\!\rightarrow \! 0$, $T\!\rightarrow \!\infty$
and $T\!\rightarrow \!\infty$, respectively. Important issues of
renormalizibility of composite operators at finite temperature are
addressed and the improved energy--momentum tensor is constructed.
The utility of the hydrostatic pressure in the non--equilibrium
quantum systems is discussed.
\\

\vspace{3mm}
\end{abstract}
\vskip5mm

\renewcommand{\thefootnote}{\arabic{footnote}}
\setcounter{footnote}{0}

\typeout{--- Main Text Start ---}

\section{Introduction}

In order to give a theoretical description of the properties of
matter under extreme conditions (like neutron stars, the early
universe or heavy--ion collisions) one is often forced to use the
statistical quantum field theory (QFT). The latter is due to
inherent quantum nature of these processes and due to an
overwhelming number of degrees of freedom involved. In recent
years, considerable effort has been devoted to the understanding
of both equilibrium and non--equilibrium behavior of such systems
(see e.g.\tsecite{Mot,Mot2} and citations therein). In fact, the
equilibrium description is worked out relatively well and number
of methodologies for doing quantum field theory on systems at or
near (local) equilibrium is available. On this level two modes of
description have been formulated: imaginary--time (or Matsubara)
approach\tsecite{LW,LB,AD,Kap1} and real--time
approach\tsecite{LW,LB,AD,TA}. In contrast to equilibrium, the
theoretical understanding of non--equilibrium quantum field
theories is still very rudimentary. Complications involved are
essentially twofold. The first is related to an appropriate choice
of the non--equilibrium initial--time conditions and their
implementation into quantum description\tsecite{Mot2,Lich1}. The
second problem is to construct the density matrix pertinent to the
level of description one aims at. The latter requires usually some
sort of coarse--graining (e.g. truncation of higher point Wigner
functions in the infinite tower of Schwinger--Dyson
equations\tsecite{Kal1}) or projecting over irrelevant subsystems
(incorporated e.g. via projection operator method\tsecite{Zw1} or
maximal entropy - MaXent - prescription\tsecite{PJ2}). However,
when the density matrix is known one may, in principle, apply the
cummulant expansion to convert the calculations into those
mimicking usual equilibrium techniques\tsecite{Kal1,KCC}. Yet, the
boundary problem prohibits {\em per se} many of equilibrium
approaches. Imaginary--time approach is clearly not applicable due
to its lack of the explicit time dependence and build--in
equilibrium (Kubo--Martin--Schwinger) boundary conditions. Among
the real--time formalisms only the Schwinger--Keldysh or
closed--time path formalism (CTP)\tsecite{LW,LB,Kal1,Coop1,EP1}
and thermo field dynamics (TFD)\tsecite{LW,LMU1,he2} has found a
wider utility in non--equilibrium computations. The CTP
formulation was conveniently applied, for instance, in the study
of non--equilibrium gluon matter\tsecite{Greiner}, cosmological
back reaction problem\tsecite{Hu2} or in the time evolution of a
non--equilibrium chiral phase transition\tsecite{coop2}. On the
other hand the non--equilibrium TFD was recently used in deriving
the transport equations for dense quantum systems\tsecite{kobrin},
or in a study of transport properties of quantum fields with
continuous mass spectrum\tsecite{he2}.

\vspace{3mm}

To extract information on the underlying field dynamics or on
non--equilibrium transport characteristics one needs to specify an
appropriate set of observables (be it conductivity, damping rates,
edge temperature jumps, viscosity, etc.). Pressure is often one of
the key parameters used in the diagnostics of off--equilibrium
quantum media. Hydrostatic pressure measurements in superfluid
He$^4$ (i.e. in He~II phase)\tsecite{THNK} and in
superconductors\tsecite{DAM} provide examples. It is thus clear
that an extension of the pressure calculations to non--equilibrium
systems could enhance our predicative ability in such areas as
(realistic) phase transitions, early universe cosmology  or hot
fusion dynamics. However, the usual procedure known from
equilibrium QFT, i.e, calculations based on the partition function
or effective potential\tsecite{AD,PR,IZ,CJT} cannot be employed
here. This is because the (grand)--canonical potential from which
the {\em thermodynamic} pressure is derived does not exist away
from equilibrium. Fortunately, more general definition of
pressure, not hinging on existence of (grand)--canonical
potential, exists. This is the so called {\em hydrostatic}
pressure and its form is deduced from the expectation value of the
energy--momentum tensor. It might be shown that in thermal
equilibrium the (classical) thermodynamic and (classical)
hydrostatic pressures are identical on account of the (classical)
virial theorem\tsecite{EPap}.

\vspace{3mm}

In this and two companion papers we aim at clarifying the
calculation of the hydrostatic pressure away from equilibrium and
at studying its bearings to various non--equilibrium situations.
Calculation of the expectation value of the energy--momentum
tensor is, however, quite delicate task even in thermal
equilibrium as computations involved are qualitatively very
different from those known, for instance, from the effective
action approach. This is because the energy--momentum tensor is a
composite operator and as such it requires a different methodology
of treatment including a different approach to renormalization
issues\tsecite{LW,PB2}. It should then come as a no surprise that
in thermal QFT the equivalence between hydrostatic and
thermodynamic pressure (or effective action) is more fragile than
in corresponding classical statistical systems. In fact, the
validity of the {\em quantum} virial theorem is by no means
established conclusively, and it is conjectured that it could
break down, for instance, in gauge theories\tsecite{LW}. Besides,
there is clearly no virial theorem away from equilibrium (not even
classically) and so in such a case one must expect disparity
between hydrostatic pressure and effective action.

\vspace{3mm}

In order to understand the difficulties involved we concentrate in
the present paper on the calculation of the hydrostatic pressure
in thermal equilibrium. To this end, we utilize the CTP approach
which both in spirit and in many technical details mimics the
realistic non--equilibrium
calculations\tsecite{Kal1,PJ2,coop2,KCC}. Presented CTP formalism
in addition to its theoretical structure which is interesting in
its own right, is important because it can be with a minor changes
directly applied to translationally invariant non--equilibrium QFT
systems\tsecite{PJ2}. In order to keep the discussion as simple as
possible we illustrate our reasonings on $O(N)$ symmetric scalar
$\lambda \ \phi^4$ theory. The model is sufficiently simple yet
complex enough to serve as an illustration of basic
characteristics of the presented method in contrast to other ones
in use. The latter has the undeniable merit of being exactly
solvable in the large--$N$ limit both at zero and finite
temperature\tsecite{PR,HKl,HS,FS1,BM,F2,AKS}. It might be shown
that the leading order approximation in $1/N$ is closely related
to the Hartree--Fock mean field approximation which has been much
studied in nuclear, many--body, atomic and molecular chemistry
applications\tsecite{PR,Coop3}. In addition, in the case of a pure
state it corresponds to a Gaussian ansatz for the Schr{\"o}dinger
wave functional\tsecite{Jackiw}. We will amplify some of these
points in later papers. We should also emphasize that although the
$O(N)\; \phi^4$ theory frequently serves as a useful playground
for study of finite--temperature phase transitions with a scalar
order parameter, this point is not objective of this work and
hence we will not pursue it here.
%
%
%
%

\vspace{3mm}

The set--up of the paper is the following: In Sections \ref{PE1}
we briefly review the derivation of the thermodynamic and
hydrostatic pressures. In Section \ref{PE2} we lay down the
mathematical framework needed for the finite--temperature
renormalization of the energy--momentum tensor (for an extensive
review on renormalization of composite operators the reader may
consult e.g., refs.\tsecite{LW,Collins,Brown}). The latter is
discussed on the $O(N)\; \phi^{4}$ theory. It is a common wisdom
that the zero temperature renormalization takes care also of the
UV divergences of the corresponding finite temperature
theory\tsecite{LW,AD,EM}. The situation with energy--momentum
tensor is, however, more complicated as there is no well defined
expectation value of the stress tensor at $T=0$\tsecite{LW,PB2}.
We show how this problem can be amended at finite temperature. The
key original results obtained here is the prescription for the
improved energy--momentum tensor of the $O(N)\; \phi^{4}$ theory.
The latter is achieved by means of the Zimmerman forest formula.
With the help of the improved stress tensor we are able to find
the corresponding QFT extension of hydrostatic pressure and hence
obtain the prescription for the renormalized pressure. This latter
result is also original finding. As a byproduct we renormalize
$\phi_{a}^{2}$ and $\phi_{a}\phi_{b}$ operators.

\vspace{3mm}

Resumed form for the pressure in the large--$N$ limit, together
with the discussion of both coupling constant and mass
renormalization is presented in Section \ref{PE3}. Calculations
are substantially simplified by use of the thermal
Dyson--Schwinger equations. For simplicity's sake our analysis is
confined to the part of the parameter space where the ground state
at large $N$ has the $O(N)$ symmetry of the original Lagrangian
and the spontaneous symmetry breakdown and Goldstone phenomena are
not possible (Bardeen and Moshe's parameter space\tsecite{BM}).

\vspace{3mm}

In Section \ref{HTE} we end up with the high--temperature
expansion of the pressure. Calculations are performed in $D=4$
both for massive and massless fields, and the result is expressed
in terms of the renormalized mass $m_{r}(T)$ and the thermal mass
shift $\delta m^2(T)$. The expansion is done by means of the
Mellin transform technique. In appropriate limits we recover the
results of Drummond {\em et al.}\tsecite{ID1} and Amelino--Camelia
and Pi\tsecite{ACP} for thermodynamic pressure (effective action).

\vspace{3mm}

The paper is furnished with two appendices. In Appendix A we
clarify some mathematical manipulations needed in Section
\ref{PE3}. For the completeness'   sake    we compute in Appendix
B  the high--temperature expansion  of the thermal--mass shift
$\delta m^{2}(T)$ which will prove useful in Section \ref{HTE}.

%

\section{Hydrostatic pressure} \label{PE1}

In thermal quantum field theory where one deals with systems in
thermal equilibrium there is an easy prescription for a pressure
calculation. The latter is based on the observation that for
thermally equilibrated systems the grand canonical partition
function $Z$ is given as
\begin{equation}
Z = e^{-\beta \Omega} = Tr(e^{-\beta(H-\mu_{i}N_{i})})\, ,
\tselea{1}
\end{equation}
\noindent where $\Omega$ is the grand canonical potential, $H$ is the
Hamiltonian, $N_{i}$ are conserved charges, $\mu_{i}$ are corresponding
chemical potentials, and $\beta$ is the inverse temperature: $\beta = 1/T$
($k_{B}=1$). Using identity $\beta \frac{\partial}{\partial \beta}= -T
\frac{\partial}{\partial T}$ together with (\tseref{1}) one gets
\begin{equation}
 T \left( \frac{\partial \Omega}{\partial T} \right)_{\mu_{i},V} =
\Omega -E + \mu_{i} N_{i}\, , \tselea{2}
\end{equation}
\noindent with $E$ and $V$ being the averaged energy and volume of
the system respectively. A comparison of (\tseref{2}) with a
corresponding thermodynamic expression for the grand canonical
potential\tsecite{LW,GM,Cub,Call1} requires that entropy $S=-
\left(\frac{\partial \Omega}{\partial T} \right)_{\mu_{i}, V}$, so
that
\begin{equation}
d\Omega = -SdT - pdV -N_{i}d\mu_{i}\; \Rightarrow \; p= -\left(
\frac{\partial \Omega}{\partial V} \; \right)_{\mu_{i}, T}\, .
\tselea{3}
\end{equation}
\noindent For large systems one can usually neglect surface effects so $E$
and $N_{i}$ become extensive quantities. Eq.(\tseref{1}) then immediately
implies that $\Omega$ is extensive quantity as well, so (\tseref{3})
simplifies to
\begin{equation}
p = -\frac{\Omega}{V} = \frac{\mbox{ln}Z}{\beta V}\, . \tselea{4}
\end{equation}
\noindent The pressure defined by Eq.(\tseref{4}) is so called
thermodynamic pressure.

\vspace{3mm}

Since $\mbox{ln}Z$ can be systematically calculated summing up all
connected closed diagrams (i.e. bubble
diagrams)\tsecite{LW,LL,DJ}, the pressure calculated via
(\tseref{4}) enjoys a considerable
popularity\tsecite{ID1,LW,LB,ID}. Unfortunately, the latter
procedure can not be extended to out of equilibrium as there is,
in general, no definition of the partition function $Z$ nor
grand--canonical potential $\Omega$ away from an equilibrium.

\vspace{3mm}

Yet another, alternative definition of a pressure not hinging on
thermodynamics can be provided; namely the hydrostatic pressure
which is formulated through the energy--momentum tensor
$\Theta^{\mu \nu}$. The formal argument leading to the hydrostatic
pressure in $D$ space--time dimensions is based on the observation
that $\langle \Theta^{0 j}(x) \rangle$ is the mean (or
macroscopic) density of momenta ${\vect{p}}^{j}$ in the point
$x^{\mu}$. Let ${\vect{P}}$ be the mean total $(D-1)$--momentum of
an infinitesimal volume $V^{(D-1)}$ centered at ${\vect{x}}$, then
the rate of change of $j$--component of ${\vect{P}}$ reads
\begin{equation}-
\frac{d{\vect{P}}^{j}(x)}{dt} = \int_{V^{(D-1)}}d^{D-1}{\vect{x}}'
\; \frac{\partial}{\partial x^{0}} \langle \Theta^{0
j}(x^{0},{\vect{x}}' ) \rangle = \sum_{i=1}^{D-1}\int_{\partial
V^{(D-1)}} d{\vect{s}}^{i} \; \langle \Theta^{i j} \rangle\, .
\tselea{5}
\end{equation}
\noindent In the second equality we have exploited the continuity
equation for $\langle \Theta^{\mu j} \rangle$ and successively we
have used Gauss's theorem\footnote{The macroscopic conservation
law for $\langle \Theta^{\mu \nu} \rangle$ (i.e. the continuity
equation) has to be postulated. For some systems, however, the
later can be directly derived from the corresponding microscopic
conservation law\tsecite{DG}.}. The $\partial V^{(D-1)}$
corresponds to the surface of $V^{(D-1)}$.

\vspace{3mm}

Anticipating  a system out  of equilibrium, we must assume a
non--trivial  distribution   of   the    mean  particle
four--velocity $U^{\mu}(x)$ (hydrodynamic velocity). Now, a
pressure is by definition a scalar quantity.  This particularly
means  that it should not depend on the hydrodynamic velocity. We
must thus go to  the local rest frame and evaluate pressure there.
However, in  the local rest frame, unlike the  equilibrium, the
notion of  a  pressure  acting equally in   all directions is
lost.   In order   to retain  the scalar  character of pressure,
one customarily defines  the {\em pressure  at a point} (in the
following denoted as  $p(x)$)\tsecite{Bach}, which is  simply the
``averaged pressure" {\footnote {To  be  precise, we should  talk
about averaging the normal  components of stress\tsecite{Bach}.}}
over all directions at  a given point.  In the local rest frame
Eq.(\tseref{5}) describes $j$--component  of the force  exerted by
the medium  on the infinitesimal volume $V^{(D-1)}$.  (By
definition there is no contribution to
$\frac{d{\vect{P}}^{j}(x)}{dt}$ caused by  the particle convection
through  $V^{(D-1)}$.) Averaging the LHS  of (\tseref{5}) over all
directions of     the normal ${\vect{n}}({\vect{x}})$, we
get\footnote{The angular average  is standardly defined for
scalars (say, $A$) as; $\int A \;d\Omega({\vect{n}})/ \int
d\Omega({\vect{n}})$, and for vectors (say, ${\vect{A}}^{i}$) as;
$\sum_{j}\int {\vect{A}}^{j} \; {\vect{n}}^{j}\;      d\Omega
({\vect{n}})/      \int d\Omega({\vect{n}})$. Similarly   we might
write the angular averages for tensors of a higher rank.}
\begin{eqnarray}
\mbox{$\frac{1}{\left(S^{D-2}_{1}\right)}$}\sum_{j=1}^{D-1}\int \;
\frac{d{\vect{P}}^{j}(x)}{dt}\; {\vect{n}}^{j}
\; d
\Omega({\vect{n}}) &=&
\mbox{$\frac{1}{\left(S^{D-2}_{1}\right)}$}\sum_{j,i=1}^{D-1}\int_{\partial
V^{(D-1)}}
ds\; \langle \Theta^{ij}(x') \rangle \; \int d\Omega({\vect{n}})\;
{\vect{n}}^{i}{\vect{n}}^{j}\nonumber\\
&=& - \frac{1}{(D-1)}\sum_{i=1}^{D-1} \int_{\partial V^{(D-1)}}ds
\; \langle \Theta^{i}_{\; i}(x') \rangle\, , \tselea{EMT22}
\end{eqnarray}
\noindent where $d\Omega({\vect{n}})$ is an element of solid angle about
${\vect{n}}$ and $S^{D-2}_{1}$ is the surface of $(D-2)$-sphere with unit
radius ($\int d\Omega({\vect{n}}) = S^{D-2}_{1} = 2
\pi^{\frac{D-1}{2}}/\Gamma (\mbox{$\frac{D-1}{2}$})$) . On the other hand,
from the definition of the pressure at a point $x^{\mu}$ we might write
\begin{equation}
\left(S^{D-2}_{1}\right)^{-1}\sum_{j=1}^{D-1}\int \;
\frac{d{\vect{P}}^{j}(x)}{dt}\; {\vect{n}}^{j} \; d
\Omega({\vect{n}}) = - p(x) \; \int_{\partial V^{(D-1)}}ds\, ,
\tseleq{EMT23}
\end{equation}
\noindent here the minus sign reflects that the force responsible
for a compression (conventionally assigned as a positive pressure)
has reversed orientation than the surface normals ${\vect{n}}$
(pointing outward). In order to keep track with the standard
text--book definition of a sign of a pressure\tsecite{Cub,Bach} we
have used in (\tseref{EMT23}) the normal ${\vect{n}}$ in a
contravariant notation (note, ${\vect{n}}^{i} = -
{\vect{n}}_{i}$). Comparing (\tseref{EMT22}) with (\tseref{EMT23})
we can write for a sufficiently small volume $V^{(D-1)}$
\begin{equation}
p(x)= - \frac{1}{(D-1)} \sum_{i=1}^{D-1}\langle
\Theta^{i}_{\;i}(x) \rangle\, . \tseleq{EMT24}
\end{equation}
\noindent We should point out that in equilibrium the
thermodynamic pressure is usually identified with the hydrostatic
one via the virial theorem\tsecite{LW,Zub}. In the remainder of
this note we shall deal with the hydrostatic pressure at
equilibrium. We shall denote the foregoing as ${\cali{P}}(T)$,
where $T$ stands for temperature. We consider the non--equilibrium
case in a future paper.

\section{Renormalization} \label{PE2}

If we proceed with (\tseref{EMT24}) to QFT this leads to the
notorious difficulties connected with the fact that $\Theta^{\mu
\nu}$ is a (local) composite operator.  If only a free theory
would be in question then the normal ordering prescription would
be sufficient to render $\langle \Theta^{\mu \nu} \rangle$ finite.
In the general case, when the interacting theory is of interest,
one must work with the Zimmerman {\em normal} ordering
prescription instead. Let us demonstrate the latter on the $O(N)\;
\phi^{4}$ theory. (In this Section we keep $N$ arbitrary.) Such a
theory is defined by the bare Lagrange function
\begin{equation}
{\cali{L}}= \frac{1}{2}\sum_{a=1}^{N}\left( (\partial
\phi_{a})^{2}-m_{0}^{2}\phi_{a}^{2} \right) -
\frac{\lambda_{0}}{8N}\left( \sum_{a=1}^{N} (\phi_{a})^{2}
\right)^{2}\, , \tseleq{6}
\end{equation}
\noindent and we assume that $m^{2}_{0}>0$. The corresponding {\em
canonical} energy--momentum tensor is given by
\begin{equation}
\Theta^{\mu \nu}_{c} =
\sum_{a}\partial^{\mu}\phi_{a}\partial^{\nu}\phi_{a} - g^{\mu \nu}
{\cali{L}}\, . \tseleq{7}
\end{equation}
\noindent The Feynman rules for Green's functions with the
energy--momentum insertion can be easily explained in momentum
space. In the reasonings to follow we shall need the (thermal)
composite Green's function\footnote{By $\phi$ we shall mean the
field in the Heisenberg picture. The subscript $H$ will be
introduced in cases when a possible ambiguity could occur.}
\begin{equation}
D^{\mu \nu}(x^{n}|y) = \langle {\cali{T}}^{*}\left\{
\phi_{r}(x_{1}) \ldots \phi_{r}(x_{n}) \Theta^{\mu \nu}_{c}(y)
\right\} \rangle\, . \tseleq{8}
\end{equation}
\noindent Here the subscript $r$ denotes the renormalized fields
in the Heisenberg picture (the internal indices are suppressed)
and ${\cali{T}}^{*}$ is the so called ${\cali{T}}^{*}$ product (or
covariant ${\cali{T}}$ product)\tsecite{IZ,N,RJ,CCR}.  The
${\cali{T}}^{*}$ product is defined in such a way that it is
simply the ${\cali{T}}$ product with all differential operators
${\cali{D}}_{\mu_{i}}$ pulled out of the ${\cali{T}}$--ordering
symbol, i.e.
\begin{equation}
{\cali{T}}^{*}\{ {\cali{D}}_{\mu_{1}}^{x_{1}}\phi_{r}(x_{1})\ldots
{\cali{D}}_{\mu_{n}}^{x_{n}}\phi_{r}(x_{n}) \} =
{\cali{D}}(i\partial_{\{\mu\}}) {\cali{T}}\{\phi_{r}(x_{1})\ldots
\phi_{r}(x_{n})\}\, , \tseleq{TP1}
\end{equation}
\vspace{1mm}
\noindent where ${\cali{D}}(i\partial_{\{\mu\}})$ is just a useful
 short--hand notation for
${\cali{D}}_{\mu_{1}}^{x_{1}}{\cali{D}}_{\mu_{2}}^{x_{2}}\ldots
{\cali{D}}_{\mu_{n}}^{x_{n}}$. In the case of thermal Green's
functions, ${\cali{T}}$ represents a contour ordering
symbol\tsecite{LW,LB,AD}. It is the mean value of the
${\cali{T}}^{*}$ ordered fields rather than the ${\cali{T}}$ ones,
which corresponds at $T=0$ and at equilibrium to the Feynman path
integral representation of Green's functions\tsecite{CCR,JS}.

\vspace{3mm}

A typical contribution to $\Theta^{\mu \nu}_{c}(y)$ can be written
as
\begin{equation}
{\cali{D}}_{\mu_{1}}\phi(y) \; {\cali{D}}_{\mu_{2}}\phi(y) \ldots
{\cali{D}}_{\mu_{n}}\phi(y)\, , \tseleq{CO1}
\end{equation}
\noindent so the typical term in (\tseref{8}) is
\begin{displaymath}
{\cali{D}}(i\partial_{\{\mu\}})\; \langle {\cali{T}}^{*}\left\{
\phi_{r}(x_{1})\ldots \phi_{r}(x_{n})\phi(y_{1})\ldots \phi(y_{k})
\right\}\rangle \left. \right|_{y_{i}=y}\equiv D_{\{\mu
\}}(x^{n}|y^{k})\left. \right|_{y_{i}=y}\, .
\end{displaymath}
\noindent Performing the Fourier transform in (\tseref{8}) we get
\begin{equation}
D^{\mu \nu}(p^{n}|p)= \sum_{k=\{2,4\}}\; \int  \left(
\prod_{i=1}^{k} \frac{d^{D}q_{i}}{(2\pi)^{D}}\right) (2\pi)^{D}\;
\delta^{D}(p-\sum_{j=1}^{k}q_{j})\; {\cali{D}}_{(k)}^{\mu
\nu}(q_{\{ \mu \}}) \; D(p^{n}|q^{k})\, , \tseleq{11}
\end{equation}
\noindent where ${\cali{D}}_{(k)}^{\mu \nu}(\ldots)$ is a Fourier
transformed differential operator corresponding to the quadratic
($k=2$) and quartic ($k=4$) terms in $\Theta^{\mu \nu}_{c}$.
Denoting the new vertex corresponding to ${\cali{D}}_{(k)}^{\mu
\nu}(\ldots)$ as $\otimes$, we can graphically represent
(\tseref{8}) through (\tseref{11}) as

\begin{figure}[h]
\begin{center}
\leavevmode
\hbox{%
\epsfxsize=7cm \epsffile{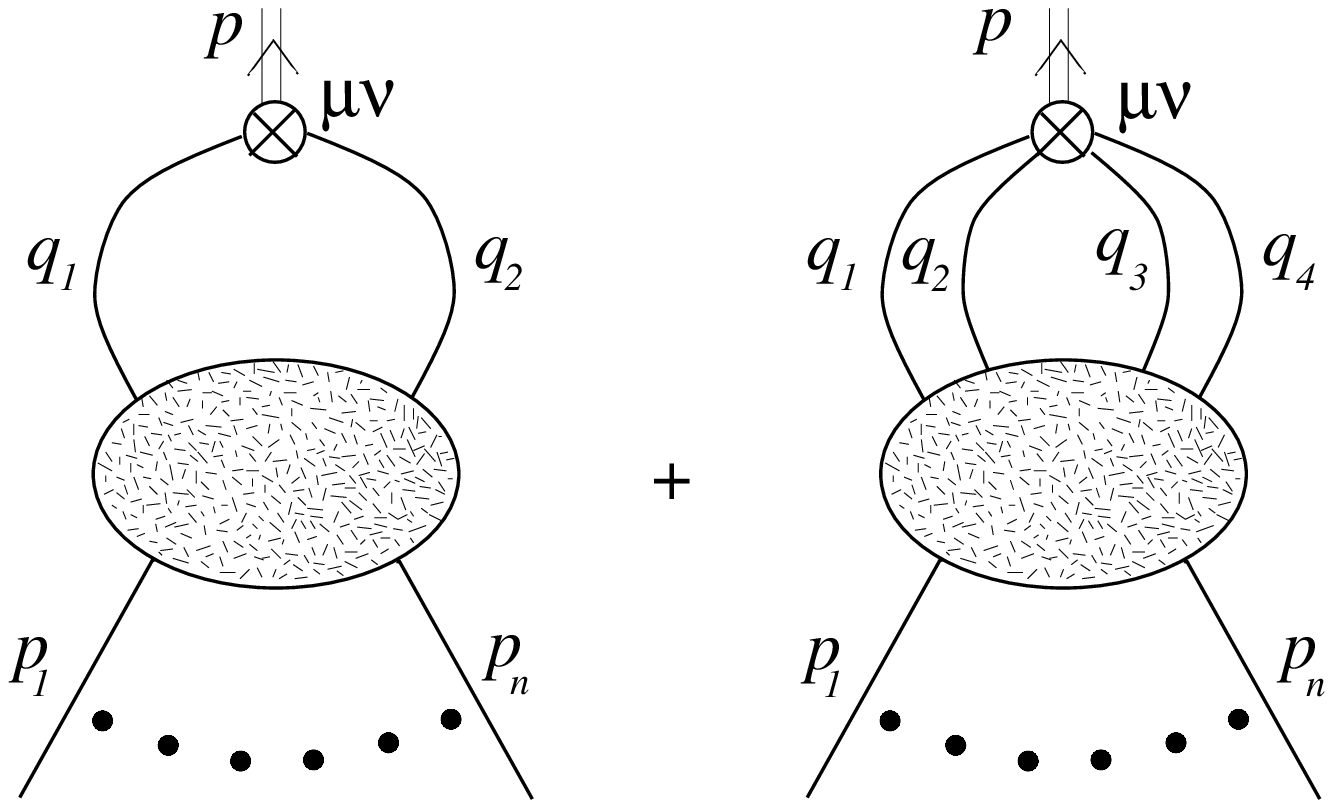}} \caption{\em The graphical
representation of $D^{\mu \nu}(p^{n}|p)$.} \label{fig1}
\end{center}
\begin{picture}(10,10)
\put(50,100){$D^{\mu \nu}(p^{n}|p)=$}
\end{picture}
\end{figure}
\vspace{3mm}
\noindent For the case at hand one can easily read off from (\tseref{7})
an explicit form of the bare composite vertices, the foregoing are

\begin{figure}[h]
\begin{flushleft}
\leavevmode
\hbox{%
\epsfxsize=2cm
\epsffile{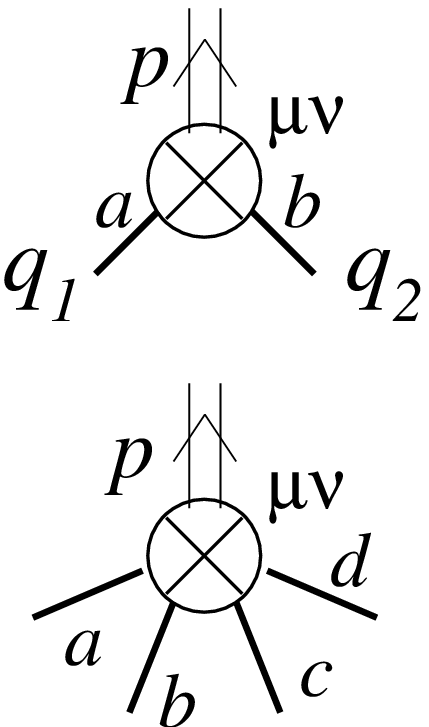}}
\end{flushleft}
\begin{picture}(20,5)
\put(65,105){$\sim \;{\cali{D}}_{(2)}^{\mu \nu}(q_{\{ \mu \}}) =
\frac{1}{2}\;\delta_{ab}\;\{ 2(q_{1}-p)^{\mu}q_{1}^{\nu}-g^{\mu
\nu}((q_{1}-p)_{\lambda}q_{1}^{\lambda}-m_{0}^{2})\}$}
\put(65,50){$\sim \;{\cali{D}}_{(4)}^{\mu \nu}(q_{\{ \mu \}}) =
\frac{g^{\mu \nu} \lambda_{0}}{8N}\{
2(\delta_{ab}\delta_{cd}+\delta_{ac}\delta_{bd}+\delta_{ad}\delta_{bc})
- 5 \delta_{ab}\delta_{cd}\delta_{ac}\}$}
\end{picture}
\end{figure}
\noindent (For the internal indices we do not adopt Einstein's
summation convention.) The blobs in Fig.\ref{fig1} comprise the
sum of all $n+2$-- and $n+4$-- (not necessarily connected) Green
functions. As usual, the disjoint bubble diagrams in Green
functions (blobs) can been divided out from the very beginning. We
have also implicitly assumed that the summation over internal
indices is understood.

\vspace{3mm}

In case when we deal with finite temperature, we choose the
contour ordering in (\ref{8}) to run along the time contour
depicted in Fig.\ref{fig18}. It is possible to show that for
Green's function calculations only horizontal paths
contribute\tsecite{EP1,LB3,FG1}. In addition, the ``physical"
fields occurring on the external lines of Green's functions have
time arguments on the upper horizontal path (type--1 fields) while
the ``ghost" fields have time arguments on the lower horizontal
path (type--2 fields). The latter modify the Feynman rules in a
nontrivial fashion\tsecite{LB,AD,EP1}. From the foregoing
discussion should be clear that in the case of thermal composite
Green's function, the new (composite) vertices are of type--1 as
the fields from which they are deduced are all
physical\footnote{For a brief introduction to the real--time
formalism in thermal QFT see for example\tsecite{LW,LB,TA}.}.

\subsection{Renormalization of $\phi_{a}(x)\phi_{b}(x)$}

Now, if there would be no $\Theta^{\mu \nu}_{c}$ insertion in
(\tseref{8}), the latter would be finite, and  so it is natural to
define the renormalized energy--momentum tensor $[\Theta^{\mu
\nu}_{c}]$ (or Zimmermann normal ordering) in such a way that
\begin{displaymath}
D_{r}^{\mu \nu}(x^{n}|y) = \langle {\cali{T}}^{*}\left\{
\phi_{r}(x_{1}) \ldots \phi_{r}(x_{n})\; [\Theta^{\mu \nu}_{c}]
\right\} \rangle\, ,
\end{displaymath}
\noindent is finite for any $n > 0$. To see what is involved, we
illustrate the mechanism of the composite operator renormalization
on $\phi_{a}(x)\phi_{b}(x)$ first - the energy--momentum tensor
case will be postponed to Section~\ref{PE22}. In the following we
shall use the mass--independent renormalization, and for
definiteness we chose the minimal subtraction scheme (MS). In MS
we can expand the bare parameters into the Laurent series which
has a simple form\tsecite{IZ,Brown,JS}, namely
\begin{equation}
\lambda_{0} = \mu^{4-D}\; \lambda_{r} \left( 1 + \sum_{k=1}^{\infty}
\frac{a_{k}(\lambda_{r}; D)}{(D-4)^{k}}\right)
\tseleq{CO2}
\end{equation}
\begin{equation}
m_{0}^{2} = m_{r}^{2} \left( 1 +
\sum_{k=1}^{\infty}\frac{b_{k}(\lambda_{r}; D)}{(D-4)^{k}}
\right)\, . \tseleq{CO3}
\end{equation}
\noindent Here $a_{0}$ and $b_{0}$ are analytic in $D=4$. The parameter
$\mu$ is the scale introduced by the renormalization in order to keep
$\lambda_{r}$ dimensionless. An important point is that both $a_{k}$'s and
$b_{k}$'s are mass, temperature and momentum independent.

\vspace{3mm}

It was Zimmermann who first realized that the forest formula known
from the ordinary Green's function renormalization\tsecite{IZ,
Collins} can be also utilized for the composite Green's functions
rendering them finite\tsecite{Collins,Zimm}. That is, we start
with Feynman diagrams expressed in terms of physical (i.e. finite)
coupling constants and masses. As we calculate diagrams to a given
order, we meet UV divergences which might be cancelled by adding
counterterm diagrams. The forest formula then prescribes how to
systematically cancel all the UV loop divergences by counterterms
to all orders. However, in contrast to the coupling constant
renormalization, the composite vertex need not to be renormalized
multiplicatively. We shall illustrate this fact in the sequel. Let
us also observe that in the lowest order (no loop) the
renormalized composite vertex equals to the bare one, and so to
that order $A=[A]$, for any composite operator $A$.

\vspace{3mm}

Now, from (\tseref{CO2}) and (\tseref{CO3}) follows that for any
function $F= F(m_{r}, \lambda_{r})$ we have \vspace{2mm}
\begin{displaymath}
\frac{\partial F}{\partial m^{2}_{r}} = \frac{\partial
m_{0}^{2}}{\partial m_{r}^{2}}\; \frac{\partial F}{\partial
m_{0}^{2}} = \frac{ m_{0}^{2}}{ m_{r}^{2}}\; \frac{\partial
F}{\partial m_{0}^{2}}\,  .
\end{displaymath}
\noindent So particularly for
\begin{displaymath}
F= D(x_{1}, \ldots , x_{n}) = \langle {\cali{T}}^{*} \{
\phi_{r}(x_{1}) \ldots \phi_{r}(x_{n}) \} \rangle\, ,
\end{displaymath}
\noindent one reads
\begin{eqnarray}
&&m^{2}_{r}\; \frac{\partial}{\partial m^{2}_{r}} D(x_{1}, \ldots , x_{n})
= m^{2}_{0}\; \frac{\partial}{\partial m^{2}_{0}} D(x_{1}, \ldots ,
x_{n})\nonumber \\
&&\mbox{\hspace{1.5cm}}= \left( - \frac{i}{2} \right) {\cali{N}}\;
\int d^{D}x\;\sum_{a=1}^{N}\; \int {\cali{D}}\phi\;
\phi_{r}(x_{1}) \ldots \phi_{r}(x_{n})\;
m_{0}^{2}\phi_{a}^{2}(x)\; \mbox{exp}(iS[\phi, T])\nonumber \\
&&\mbox{\hspace{1.5cm}}= \left( - \frac{i}{2} \right) \; \int
d^{D}x\;\sum_{a=1}^{N}\;D_{a}(x_{1}, \ldots, x_{n}|x; m_{0}^{2})\,
. \tselea{CO5}
\end{eqnarray}
\noindent Here ${\cali{N}}^{-1}$ is the standard denominator of the path
integral representation of Green's function. We should apply the
derivative also on ${\cali{N}}$ but this would produce disconnected graphs
with bubble diagrams. The former precisely cancel the very same
disconnected graphs in the first term, so we are finally left with no
bubble diagrams in (\tseref{CO5}). In the Fourier space (\tseref{CO5})
reads
\begin{equation}
m^{2}_{r}\; \frac{\partial}{\partial m^{2}_{r}} D(p_{1}, \ldots
,p_{n}) = \left( - \frac{i}{2}  \right) \sum_{a=1}^{N}\;
D_{a}(p_{1}, \ldots , p_{n}|0;m_{0}^{2})\, . \tseleq{CO6}
\end{equation}
\noindent As the LHS is finite there cannot be any pole terms on
the RHS either, and so $\sum_{a} m^{2}_{0}\phi_{a}^{2}$ is by
itself a renormalized composite operator. We see that $m_{0}^{2}$
precisely compensates the singularity of $\sum_{a=1}^{N}
\phi_{a}^{2}$.

\vspace{3mm}

Now, it is well known that any second--rank tensor (say $M_{a b}$)
can be generally decomposed into three irreducible tensors; an
antisymmetric tensor, a symmetric traceless tensor and an
invariant tensor. Let us set $M_{ab}=\phi_{a}\phi_{b}$, so the
symmetric traceless tensor $K_{a b}$ reads
\begin{equation}
K_{ab}(x) = \phi_{a}(x)\phi_{b}(x)-\delta_{a
b}/N\;\sum_{c=1}^{n}\phi^{2}_{c}(x)\, , \tseleq{CO64}
\end{equation}
\noindent whilst the invariant tensor $I_{a b}$ is
\begin{displaymath}
I_{a b}(x) = \delta_{a b}/N \sum_{c=1}^{N}\phi_{c}^{2}(x)\, .
\end{displaymath}
\noindent Because the renormalized composite operators have to preserve a
tensorial structure of the bare ones, we immediately have that
\begin{equation}
K_{a b} = A_{1}[K_{a b}]\;\;\; \mbox{and}\;\;\; I_{a b} =
A_{2}[I_{a b}]\, , \tseleq{CO66}
\end{equation}
\noindent where both $A_{1}$ and $A_{2}$ must have structure $(1 + \sum
(\mbox{poles}))$. The foregoing guarantees that to the lowest order $K_{a
b} = [K_{a b}]$ and $I_{a b}= [I_{a b}]$. As we saw in (\tseref{CO6}),
$m_{0}^{2}I_{a b}$ is renormalized, and so from (\tseref{CO66}) follows
that $m_{0}^{2}I_{a b} = C\; [I_{a b}]$. Here $C$ has dimension $[m^{2}]$
and is analytic in $D=4$.  We can uniquely set $C=m^{2}_{r}$ because only
this choice fulfils the lowest order condition $I_{ab}=[I_{ab}]$ (c.f.
Eq.(\tseref{CO3})). Collecting our results together we might write
\begin{equation}
\sum_{c} \phi_{c}^{2} = Z_{\Sigma \phi^{2}}\; \left[ \sum_{c}
\phi_{c}^{2} \right] = Z_{\Sigma \phi^{2}} \sum_{c}
[\phi_{c}^{2}]\, , \tseleq{CO67}
\end{equation}
\noindent with $Z_{\Sigma\phi^{2}} = A_{2}=
\frac{m^{2}{r}}{m^{2}_{0}}$. In the second equality we have used
an obvious linearity\tsecite{Collins} of $[\ldots]$. From
(\tseref{CO64}) and (\tseref{CO67}) follows that
\begin{equation}
\phi_{a}(x)\phi_{b}(x) = A_{1}[\phi_{a}(x)\phi_{b}(x)] -
\frac{\delta_{ab}}{N}(A_{1} - Z_{\Sigma \phi^{2}})\;
\sum_{c=1}^{N}[\phi_{c}^{2}(x)]\, . \tseleq{CO57}
\end{equation}
\noindent So particularly for $\phi^{2}_{a}$ one reads
\begin{equation}
\phi^{2}_{a} = \frac{1}{N}\left( (N-1)A_{1} + Z_{\Sigma \phi^{2}}
\right) \; [\phi^{2}_{a}] - \frac{1}{N} \left( A_{1} - Z_{\Sigma
\phi^{2}} \right) \sum_{c \not = a } [\phi_{c}^{2}]\, .
\tseleq{CO58}
\end{equation}
\noindent From the discussion above it does not seem to be
possible to obtain more information about $A_{1}$ without doing an
explicit perturbative calculations, however it is easy to
demonstrate that $A_{1} \not= Z_{\Sigma \phi^{2}}$. To show this,
let us consider the simplest non--trivial case; i.e. N=2, and
calculate $A_{1}$ to order $\lambda_{r}$. For that we need to
discuss the renormalization of the $n$--point composite Green's
function with, say, $\phi^{2}_{1}$ insertion. To do that, it
suffices to discuss the renormalization of the corresponding 1PI
$n$-point Green's function. The perturbative expansion for the
composite vertex to order $\lambda_{r}$ can be easily generated
via the Dyson--Schwinger (DS) equation \tsecite{PC} and it
reads\footnote{Throughout the paper we accept the usual
convention: Ordinary (not necessarily connected) N-point Green's
functions are represented with dotted blobs with N external legs,
connected N-point Green's functions are represented with hatched
blobs with N external legs and 1-PI  N-point Green's functions are
represented with cross hatched blobs with N truncated legs
(represented by solid circles in vertices). }

\vspace{3cm}

\begin{figure}[h]
\begin{center}
\leavevmode
\hbox{%
\epsfxsize=12.7cm \epsffile{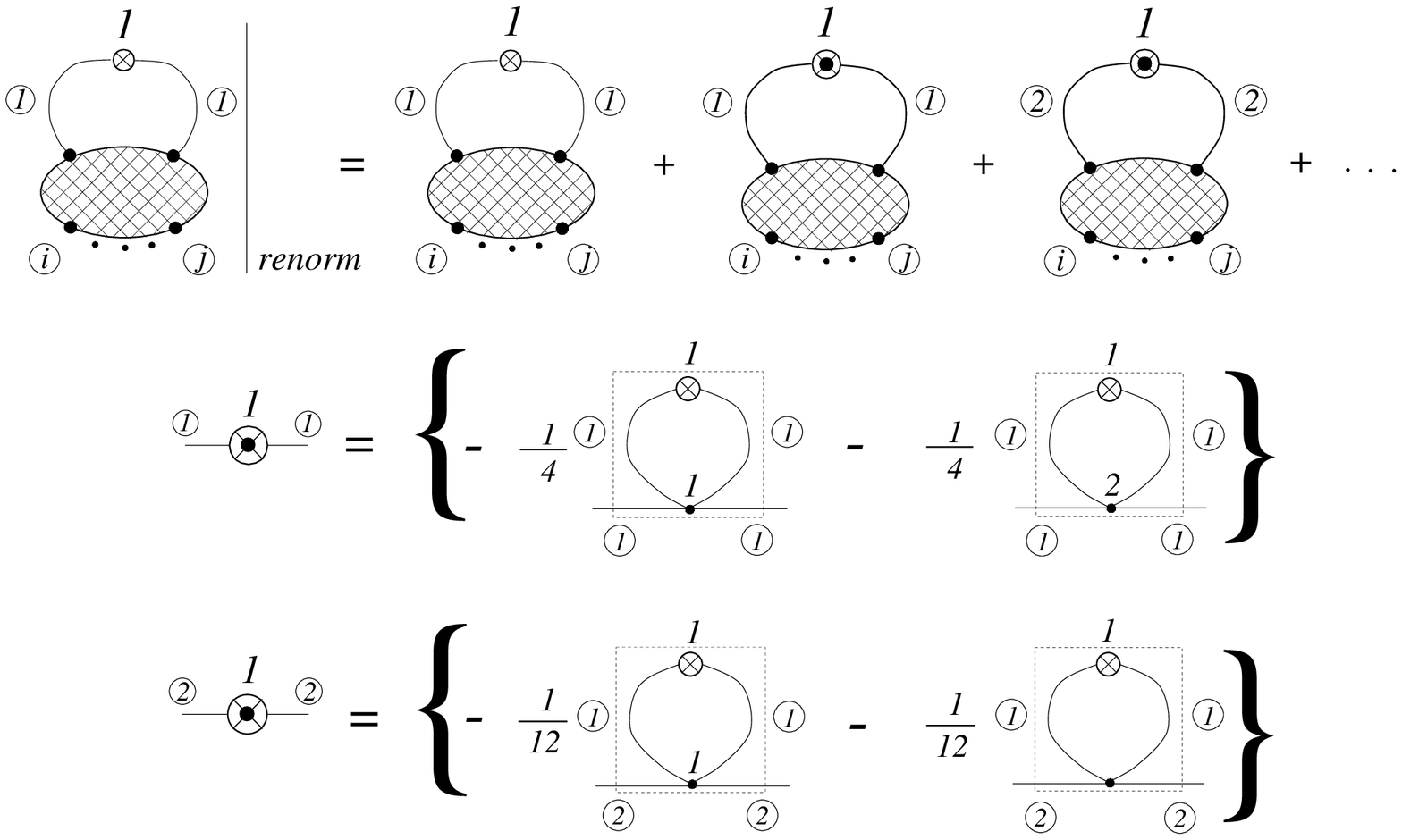}}
\end{center}
\setlength{\unitlength}{1mm}
\begin{picture}(20,7)
\put(0,47){where} \put(139,72){(3.16)}
\end{picture}
\end{figure}
\addtocounter{equation}{1}
\noindent Here cross--hatched blobs refer to (renormalized) 1PI
$(n+2)$--point Green's function, circled indices mark a type of
the field propagated on the indicated line, and uncircled numbers
refer to thermal indices (we explicitly indicate only relevant
thermal indices). The counterterms, symbolized by a heavy dot, are
extracted from the boxed diagrams (elementary Zimmermann forests).
In MS scheme one gets the following results:

\begin{figure}[h]
\begin{flushleft}
\leavevmode
\hbox{%
\epsfxsize=2cm
\epsffile{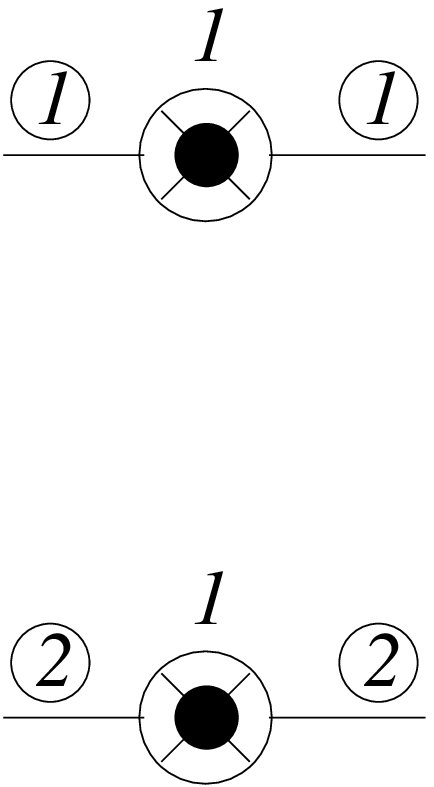}}
\end{flushleft}
\setlength{\unitlength}{1mm}
\begin{picture}(15,10)
\put(25,43){$= \frac{i\;\lambda_{r}\mu^{4-D}}{4}\int
\frac{d^{D}q}{(2pi)^{D}} \left\{ \D_{11}(q)\D_{11}(-q) -
\D_{12}(q)\D_{12}(-q) \right\}|_{\mbox{{\scriptsize MS pole term}}}$}
\put(25,32){$= - \frac{1}{4} \partial_{m_{r}^{2}}\left(
\frac{\Gamma\left(1-\frac{D}{2}\right)}{(4 \pi)^{\frac{D}{2}}} \;
\lambda_{r}\; \mu^{4-D}\; m_{r}^{D-2}\right)|_{\mbox{\scriptsize MS}} =
- \lambda_{r} \mu^{4-D} /2\; (D-4)\; (4 \pi)^{2}$}
\put(25,17){$ = - \lambda_{r} \mu^{4-D} / 6\; (D-4)\; (6 \pi)^{2}$.}
\end{picture}
\end{figure}
\vspace{-1cm}
\noindent Here $\D_{11}$ and $\D_{12}$ are the usual thermal
propagators in the real--time formalism\tsecite{LW,LB,TA} (see
also Section \ref{PE3}). From (3.16) we can directly read off that
\begin{displaymath}
[\phi^{2}_{1}] = \left(1 - \frac{\lambda_{r} \mu^{4-D}}{2\;(D-4)\; (4
\pi)^{2}} +
{\cali{O}}(\lambda_{r}^{2})\right)\;\phi_{1}^{2} + \left(-
\frac{\lambda_{r} \mu^{4-D}}{6\; (D-4)\; (4 \pi)^{2}} +
{\cali{O}}(\lambda_{r}^{2})\right)\;\phi_{2}^{2}.
\end{displaymath}
\noindent As the coefficient before $\phi_{2}^{2}$ is not zero, we
conclude that $A_{1} \not= Z_{\Sigma \phi^{2}}$. It is not a great
challenge to repeat the previous calculations for the $\phi_{1}\phi_{2}$
insertion. The latter gives
\begin{displaymath}
 A_{1} = 1 - \frac{\lambda_{r} \mu^{4-D}}{3\; (D-4)\;
(4 \pi)^{2}} +
{\cali{O}}(\lambda_{r}^{2}).
\end{displaymath}
\noindent Eq.(\tseref{CO58}) exhibits the so called operator
mixing\tsecite{IZ}; the renormalization of $\phi_{a}^{2}$ cannot
be considered independently of the renormalization of
$\phi_{c}^{2}$ ($c \not= a$). The latter is a general feature of
composite operator renormalization. Note, however, that
$\phi_{a}\phi_{b}$ ($a \not=b$) do not mix by renormalization,
i.e. they renormalize multiplicatively. It can be shown that
composite operators mix under renormalization only with those
composite operators which have dimension less or
equal\tsecite{IZ,Collins,Zimm}.

\vspace{3mm}

Unfortunately, if we apply the previous arguments to $n=0$, the
result is not finite; another additional renormalization must be
performed. The fact that the expectation values of $[\ldots]$ are
generally UV divergent, in spite of being finite for the composite
Green's functions\footnote{Also called the matrix elements of
$[\ldots]$.}, can be nicely illustrated with the composite
operator $[\phi^{2}]$ in the $N=1$ theory. Taking the diagrams for
$D(0|0)$ and applying successively the (unrenormalized) DS
equation\tsecite{LW,PC} we get

\begin{figure}[h]
\begin{center}
\leavevmode
\hbox{%
\epsfxsize=12cm \epsffile{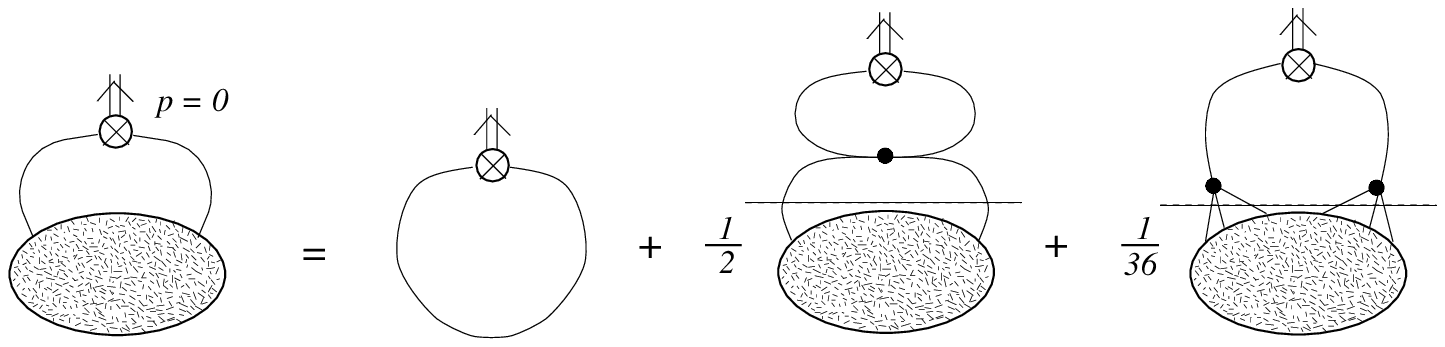}}
\end{center}
\setlength{\unitlength}{1mm}
\begin{picture}(20,7)
\put(140,20){(3.17)}
\end{picture}
\end{figure}
\addtocounter{equation}{1}
\noindent Eq.(3.17) might be rewritten as
\begin{eqnarray}
D(0|0) &=& D(0|0)|_{\lambda_{r}^{0}}\nonumber \\
&& + \frac{1}{2} \; \int
\frac{d^{D}q_{1}}{(2\pi)^{D}}\frac{d^{D}q_{2}}{(2\pi)^{D}}\;
\delta^{D}(q_{1}+ q_{2})\;
D^{amp}(q^{2}|0)|_{\lambda_{r}}\; D(q^{2})\nonumber \\
&& + \frac{1}{36} \; \int \prod_{i=1}^{6}
\frac{d^{D}q_{i}}{(2\pi)^{D}}\; \delta^{D}(\sum_{j=1}^{6}q_{j})\;
D^{amp}(q^{6}|0)|_{\lambda_{r}^{2}}\; D(q^{6})\, , \tselea{CO8}
\end{eqnarray}
\noindent where $D^{amp}(q^{m}|0)|_{\lambda_{r}^{k}}$ is the
$m$--point amputated composite Green's function to order
$\lambda_{r}^{k}$, and $D(q^{m})$ is the full $m$--point Green's
function.  The crucial point is that we can write $D(0|0)$ as a
sum of terms, which, apart from the first (free field) diagram,
are factorized to the product of the composite Green's function
with $n > 0$ and the full Green's function. (The factorization is
represented in (3.17) by the dashed lines.) Note that the
expansion (3.17) is not unique as various other ways of pulling
vertices out of Green's function may be utilized but this
particular form will prove to be important in the next section
(see Eq.(3.26)).

\vspace{3mm}

\begin{figure}[h]
\begin{center}
\leavevmode
\hbox{%
\epsfxsize=12cm \epsffile{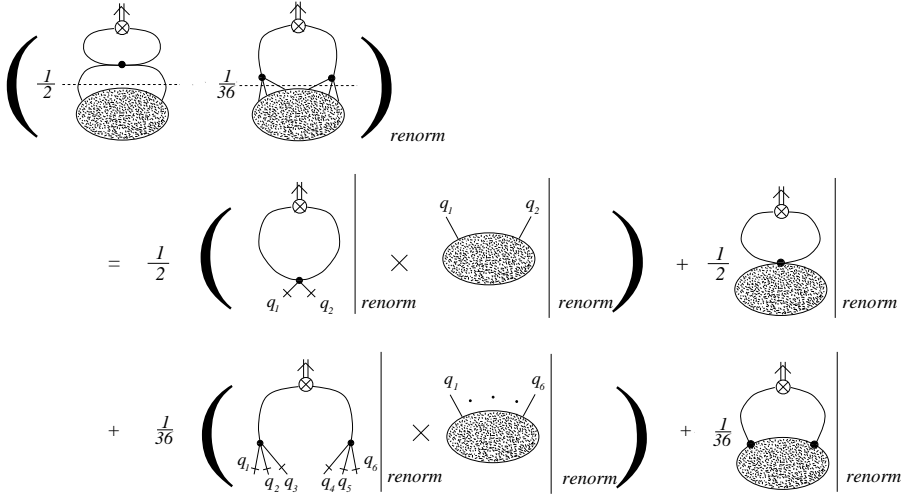}}
\caption{\em Counterterm renormalization of the last two diagrams
in Eq.(3.17). (Cut legs indicate amputations.)} \label{fig11}
\end{center}
\setlength{\unitlength}{1mm}
\begin{picture}(10,7)
\end{picture}
\end{figure}

Now, utilizing the counterterm renormalization to the last two
diagrams in (3.17) we get situation depicted in Fig.\ref{fig11}.
Terms inside of the parentheses are finite, this is because both
the composite Green's functions ($n \ge 2$ !) and the full Green's
functions are finite after renormalization. The counterterm
diagrams, which appear on the RHS of the parentheses, precisely
cancel the UV divergences coming from the loop integrations over
momenta $q_{1}\ldots q_{i}$ which must be finally performed.  The
heavy dots schematically indicates the corresponding counterterms.
In the spirit of the counterterm renormalization we should finally
subtract the counterterm associated with the overall superficial
divergence\footnote{A simple power counting in the $\phi^{4}$
theory reveals\tsecite{IZ} that for a composite operator $A$ with
dimension $\omega_{A}$ the superficial degree of divergence
$\omega$ corresponding to an $n$--point diagram is $\omega =
\omega_{A}-n$.} related to the diagrams in question. But as we saw
this is not necessary; individual counterterm diagrams (Zimmermann
forests) mutually cancel their divergences leaving behind a finite
result.

So the only UV divergence in Eq.(3.17) which cannot be cured by
existing counterterms is that coming from the first (i.e. free
field or ring) diagram. The foregoing divergence is evidently
temperature independent (to see that, simply use an explicit form
of the free thermal propagator $\D_{11}$). Hence, if we define
\begin{equation}
\langle \phi^{2} \rangle_{\mbox{\footnotesize{renorm}}} = \langle [\phi^{2}]
\rangle - \langle 0| [\phi^{2}] |0 \rangle,
\tseleq{Norm}
\end{equation}
\noindent or, alternatively
\begin{equation}
\langle \phi^{2} \rangle_{\mbox{\footnotesize{renorm}}} = \langle
[\phi^{2}] \rangle - \langle [\phi^{2}]
\rangle|_{\mbox{\footnotesize{free fields}}}\, , \tseleq{Norm2}
\end{equation}
\noindent we get  finite quantities, as desired. On the other hand, we
should emphasize that
\begin{equation}
\langle \phi^{2} \rangle - \langle 0| \phi^{2} |0 \rangle =
Z_{\phi^{2}}\left\{ \langle [\phi^{2}] \rangle - \langle 0|
[\phi^{2}] |0 \rangle \right\} \not= \mbox{finite in $D$=4}\, .
\end{equation}
\noindent An extension of the previous reasonings to any $N>1$ is
straightforward, only difference is that we must deal with operator
mixing which makes (\tseref{Norm}) and (\tseref{Norm2}) less trivial.

\vspace{3mm}

The important lesson which we have learnt here is that the naive
``double dotted'' normal product (i.e. subtraction of the vacuum
expectation value from a given operator) does not generally give a
finite result. The former is perfectly suited for the free theory
($Z_{\Sigma \phi^{2}} = 1$) but in the interacting case we must
resort to the prescription (\tseref{Norm}) or (\tseref{Norm2})
instead.

\subsection{ Renormalization of the energy--momentum
tensor}\label{PE22}

In order to calculate the hydrostatic pressure, we need to find
such $\langle \Theta^{\mu \nu}_{c}
\rangle|_{\mbox{\footnotesize{renorm}}}$ which apart from being
finite is also consistent with our derivation of the hydrostatic
pressure introduced in the introductory Section. In view of the
previous treatment, we however cannot, however, expect that
$\Theta^{\mu \nu}_{c}$ will be renormalized multiplicatively.
Instead, new terms with a different structure than $\Theta^{\mu
\nu}_{c}$ itself will be generated during renormalization. The
latter must add up to $\Theta^{\mu \nu}_{c}$ in order to render
$D^{\mu \nu}(x^{n}|y)$ finite\footnote{In fact it can be
shown\tsecite{LW,Collins} that the Noether currents corresponding
to a given internal symmetry are renormalized, i.e. $J^{a}=
[J^{a}]$, however, this is not the case for the Noether currents
corresponding to external symmetries (like $\Theta^{\mu \nu}_{c}$
is).}.

\vspace{2mm}

Now, the key ingredient exploited in Eq.(\tseref{5}) is the
conservation law (continuity equation). It is well known that one
can ``modify" $\Theta^{\mu \nu}_{C}$ in such a way that the new
tensor $\Theta^{\mu \nu}$ preserves the convergence properties of
$\Theta^{\mu \nu}_{c}$. Such a modification (the Pauli
transformation) reads
\begin{eqnarray}
&&\Theta^{\mu \nu} = \Theta^{\mu \nu}_{c} +
\partial_{\lambda}X^{\lambda \mu \nu}\, ,\nonumber \\
&& X^{\lambda \mu \nu} = -X^{\mu \lambda \nu}\, . \tselea{pp1}
\end{eqnarray}
\noindent For scalar fields (\tseref{pp1}) is the only
transformation which neither changes the divergence properties of
$\Theta^{\mu \nu}_{c}$ nor the generators of the Poincare group
constructed out of $\Theta^{\mu \nu}_{c}$\tsecite{LW,IZ,DG,RJ}.
Because the renormalized (or improved) energy momentum tensor must
be conserved (otherwise theory would be anomalous), it has to mix
with $\Theta^{\mu \nu}_{c}$ under renormalization only via the
Pauli transformation, i.e.
\begin{equation}
[\Theta^{\mu \nu}_{c}] = \Theta^{\mu \nu}_{c} +
\partial_{\lambda}X^{\lambda \mu \nu}\nonumber\, .
\tseleq{ppp1}
\end{equation}
\noindent In order to determine $X^{\lambda \mu \nu}$, we should
realize that its role is to cancel divergences present in
$\Theta^{\mu \nu}_{c}$.  Such a cancellation can be, however,
performed only by means of composite operators which are even in
the number of fields (note that $\Theta^{\mu \nu}_{c}$ is even in
fields and Green's functions with the odd number of fields
vanish).  Recalling the condition that renormalization can mix
only operators with dimension less or equal, we see that the
dimension of $X^{\lambda \mu \nu}$ must be $D-1$, and that
$X^{\lambda \mu \nu}$ must be quadratic in fields. The only
possible form which is compatible with tensorial  structure
(\tseref{pp1}) is then
\begin{equation}
X^{\lambda \mu \nu} = \sum_{a, b = 1}^{N} c_{a b}(\lambda_{r};D)\; \left(
\partial^{\mu} g^{\lambda \nu} - \partial^{\lambda} g^{\mu \nu}
\right) \; \phi_{a}\phi_{b}\, . \tseleq{ppp5}
\end{equation}
\noindent From the fact that $\Theta^{\mu \nu}_{c}$ and
$[\Theta^{\mu \nu}_{c}]$ are  $O(N)$ invariant (see
Eq.(\tseref{7})), $\partial _{\lambda} X^{\lambda \mu \nu}$ must
be also $O(N)$ invariant, so $c_{a b} = \delta_{ab} c$. Thus,
finally we can write
\begin{equation}
[\Theta^{\mu\nu}_{c}] = \Theta^{\mu \nu}_{c} + c(\lambda_{r};D)\;
\sum_{a =1}^{N} \left( \partial^{\mu}\partial^{\nu} - g^{\mu \nu}
\partial^{2}\right) \; \phi_{a}^{2}\, , \tseleq{ppp6}
\end{equation}
\noindent with $c = c_{0} + \sum (\mbox{poles})$, here $c_{0}$ is
analytic in $D$. Structure of $c(\lambda_{r};D)$ could be further
determined, similarly as in the $N=1$ theory, employing a
renormalization group equation\tsecite{Brown}. We do not intend to
do that as the detailed structure of $c$ will show totally
irrelevant for the following discussion, however, it turns out to
be important in non--equilibrium cases.

\vspace{2mm}

Now, similarly as before, $[\Theta^{\mu \nu}_{c}]$ gives the
finite composite Green's functions if $n >0$ but the expectation
value $\langle [\Theta^{\mu \nu}_{c}] \rangle$ is divergent (for
discussion of the $N=1$ theory see, e.g., Brown\tsecite{Brown}).
The unrenormalized DS equation for $D^{\mu \nu}(0|0)$
reads~\cite{LW}

\begin{figure}[h]
\begin{center}
\leavevmode
\hbox{%
\epsfxsize=13cm \epsffile{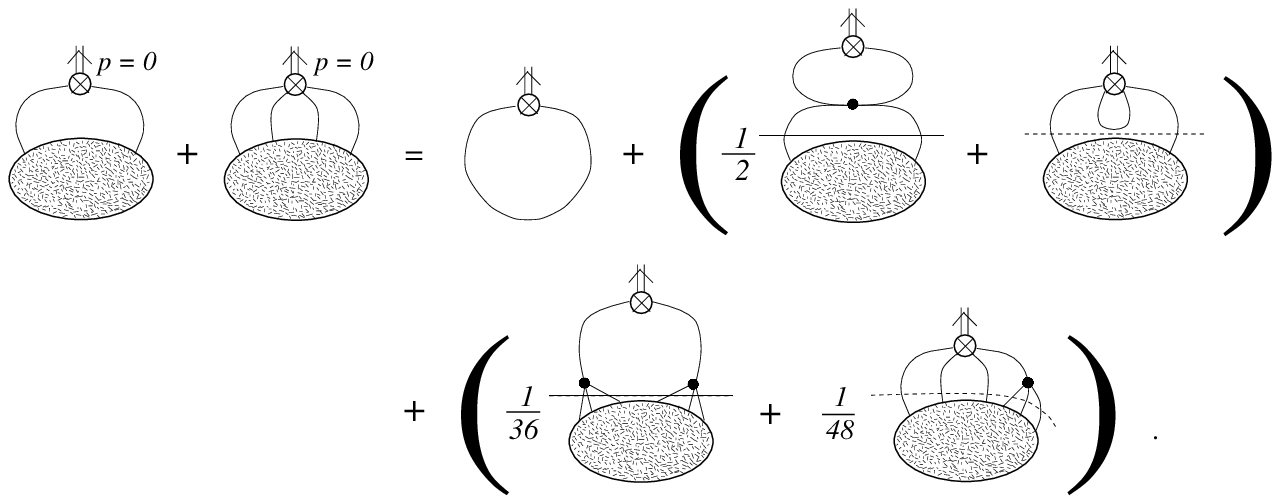}}
\end{center}
\setlength{\unitlength}{1mm}
\begin{picture}(20,7)
\put(139,18){(3.26)}
\end{picture}
\end{figure}
\addtocounter{equation}{1}
\noindent The structure of the composite vertices in (3.26) is
that described at the beginning of this Section. Note that the
amputated composite Green's functions in individual parentheses
are of the same order in $\lambda_{r}$. Performing the counterterm
renormalization as in the case of $\langle [\phi^{2}] \rangle$, we
factorize the graphs inside of parentheses into the product of the
renormalized 2-- (and 6--) point composite Green's function and
the renormalized full 2-- (and 6--) point Green's function. The
latter are finite. The UV divergences arisen during the
integrations over momenta connecting both composite and full
Green's functions are precisely cancelled by the remaining
counterterm diagrams. Only divergence comes from the free--field
contribution, more precisely from the $T=0$ ring diagram. Defining
\begin{equation} \langle \Theta^{\mu \nu}_{c} \rangle
|_{\mbox{\footnotesize{renorm}}} = \langle [\Theta^{\mu \nu}_{c}]
\rangle - \langle 0| [\Theta^{\mu \nu}_{c}]|0 \rangle,
\tseleq{ppp2}
\end{equation} \noindent or \begin{equation} \langle \Theta^{\mu \nu}_{c}
\rangle |_{\mbox{\footnotesize{renorm}}} = \langle [\Theta^{\mu
\nu}_{c}] \rangle - \langle [\Theta^{\mu \nu}_{c}] \rangle
|_{\mbox{\footnotesize{free field}}}\, , \tseleq{ppp3}
\end{equation}
\noindent we get the finite expressions. Note that the
conservation law is manifest in both cases. In equilibrium (and in
$T=0$) we can, due to space--time translational invariance of
$\langle \ldots \rangle$, write
\begin{equation}
\langle [\Theta^{\mu \nu}_{c}] \rangle = \langle \Theta^{\mu
\nu}_{c}\rangle +
\partial_{\lambda} \langle X^{\lambda \mu \nu } \rangle = \langle
\Theta^{\mu \nu}_{c}\rangle\, . \tseleq{ppp8}
\end{equation}
\noindent Using (\tseref{ppp2}) or (\tseref{ppp3}) we get either  the
thermal interaction pressure or the interaction pressure,
respectively. This can be explicitly written as
\begin{equation}
{\cali{P}}_{\mbox{\footnotesize{th.int.}}}(T) = {\cali{P}}(T) -
{\cali{P}}(0) = -\frac{1}{(D-1)} \sum_{i=1}^{D-1}\left\{ \langle
\Theta_{c\; i}^{i} \rangle - \langle 0| \Theta_{c\;i}^{i} | 0
\rangle \right\}\, , \tseleq{ppp9}
\end{equation}
\noindent or
\begin{equation}
{\cali{P}}_{\mbox{\footnotesize{int.}}}(T) = {\cali{P}}(T) -
{\cali{P}}_{\mbox{\footnotesize{free field}}}(T) =
-\frac{1}{(D-1)} \sum_{i=1}^{D-1} \left\{ \langle \Theta_{c\;
i}^{i} \rangle - \langle \Theta_{c\; i}^{i}
\rangle|_{\mbox{\footnotesize{free field}}}
 \right\}.
\tseleq{ppp10}
\end{equation}
\noindent In order to keep connection with calculations done by
Drummond {\em et al.} in\tsecite{ID1} we shall in the sequel deal
with the thermal interaction pressure only. If instead of an
equilibrium, a non--equilibrium medium would be in question,
translational invariance of $\langle \ldots \rangle$ might be
lost, in that case either prescription (\tseref{ppp2}) or
(\tseref{ppp3}) is obligatory, and consequently $c(\lambda_{r};D)$
in (\tseref{ppp6}) must be further specified.

\section{Hydrostatic pressure - calculation} \label{PE3}

In the previous section we have prepared ground for a hydrostatic
pressure calculations. In this section we aim to apply the
previous results to the massive $O(N)\; \phi^{4}$ theory in the
large--$N$ limit. Anticipating an out of equilibrium application,
we shall use the real--time formalism even if the imaginary--time
one is more natural in the equilibrium context. As we aim to
evaluate the hydrostatic pressure in 4 dimensions, we use here,
similarly as in the previous Section, the usual dimensional
regularization to regulate the theory (i.e. here and throughout we
keep $D$ slightly away from the physical value $D=4$).

\vspace{2mm}

In order to actually pursue the pressure calculation we feel it is
necessary to briefly review the mass and coupling renormalization
of the model at hand. This will also help to clarify the notation
used. While we hope to provide all essentials requisite for our
task, good discussion of alternative approaches and
renormalization prescriptions may be obtained for instance
in\tsecite{HS,FS1,AKS}.

\subsection{Mass renormalization} \label{PE31}

In the Dyson multiplicative renormalization the fact that the
complete propagator has a pole at the physical mass leads to the
usual mass renormalization prescription\tsecite{IZ}:
\begin{equation}
m_{r}^{2}= m_{0}^{2} + \Sigma(m^{2}_{r})\, , \tseleq{c1}
\end{equation}
\noindent where $m_{r}$ is renormalized mass and
$\Sigma(m_{r}^{2})$ is the proper self--energy evaluated at the
mass shell; $p^{2}= m_{r}^{2}$. In fact, Eq.(\tseref{c1}) is
nothing but the statement that 2--point vertex function
$\Gamma^{(2)}_{r}$ evaluated at the mass-shell must vanish. The
Dyson--Schwinger equation corresponding to the proper self--energy
reads\tsecite{HS,EM,PC,PJ}:
\begin{figure}[h]
\begin{center}
\leavevmode
\hbox{%
\epsfxsize=10.5cm \epsffile{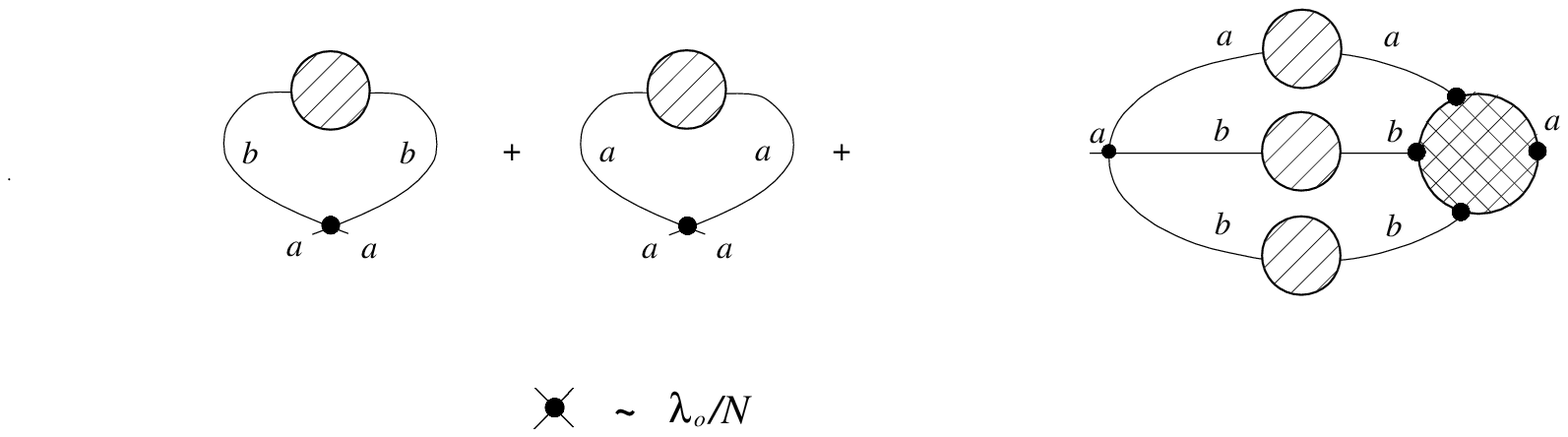}} \label{fig5}
\end{center}
\setlength{\unitlength}{1mm}
\begin{picture}(10,7)
\put(10,30){$\Sigma^{aa} =$}
\put(23.5,30){$\frac{1}{2}\sum_{b=1}^{N}$}
\put(82,30){$\frac{i}{2}\sum_{b=1}^{N}$}
\put(10,12){$\Sigma^{ac}|_{a\not= c}\; = 0\; \; \; ;$}
\put(142,30){(4.2)}
\end{picture}
\end{figure}
\vspace{-5mm}
\addtocounter{equation}{1}
\noindent where hatched blobs represent 2--point connected Green's
functions whilst cross--hatched blobs represent proper vertices
$\Gamma^{(4)}_{r}$ (i.e. 1PI 4--point Green's function).  As
$\Sigma^{aa}$ are the same for all $a$, we shall simplify notation
and write $\Sigma$ instead. In the sequel the following convention
is accepted:
\begin{figure}[h]
\begin{center}
\leavevmode
\hbox{%
\epsfxsize=2.2cm \epsffile{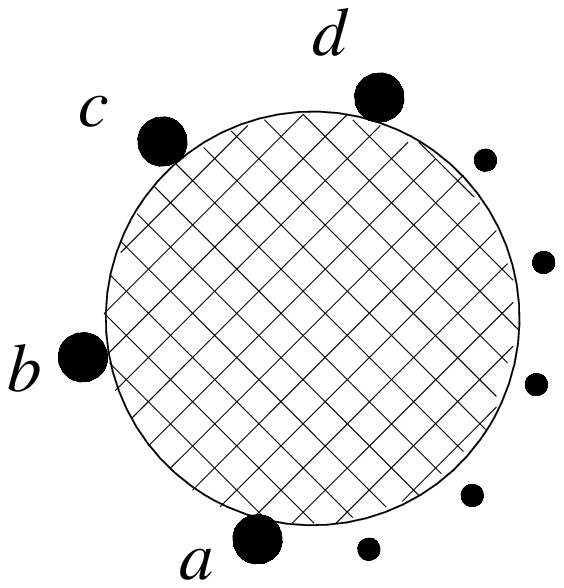}}
\end{center}
\setlength{\unitlength}{1mm}
\begin{picture}(10,7)
\put(93,21){$ =\; \Gamma^{(n)\; abcd \ldots}_{r}$}
\end{picture}
\end{figure}

\noindent The second term in (4.2) actually does not
contribute in the large--$N$ limit. It is easy to see that the
third term does not contribute either. This is because each
hatched blob behaves at most as $N^{0}$ whilst $\Gamma^{(4)}$ goes
maximally\footnote{ In the $\phi^{4}$ theory there is a simple
relation between the number of loops ($L$), vertices ($V$) and
external lines ($E$); $4V = 2I +E$. Together with the Euler
relation for connected graphs; $L= I -V +1$ (here $I$ is the
number of internal lines), we have $L-V = \frac{2-E}{2}$. As each
loop carries maximally a factor of $N$ (this is saturated only for
``tadpole" loops) and each vertex carries a factor of $N^{-1}$,
the overall blob contribution behaves at most as $N^{L-V} =
N^{\frac{2-E}{2}}.$} as $N^{-1}$. Consequently, various
contributions from the first graph in (4.2) contribute at most
$N^{0}$, whereas in the second graph the contributions contribute
up to order $N^{-1}$. So the first diagram dominates, provided we
retain only such 2--point connected Green's functions which are
proportional to $N^{0}$ (as mentioned in the footnote, these are
comprised only of tadpole loops.). After neglecting the ``setting
sun" graph, Eg.(4.2) generates upon iterating the so called
superdaisy diagrams\tsecite{ID1,CJT,EM}.

\vspace{3mm}

Let us now define $\Sigma(m^{2}_{r}) = \lambda_{0}\;
{\cali{M}}(m_{r}^{2})$. Because the tadpole diagram in (4.2) can
be easily resumed we observe that
\begin{equation}
{\cali{M}}(m_{r}^{2}) = \frac{1}{2}\int \frac{d^{D}q}{(2 \pi)^{D}}
\; \frac{i}{q^{2}-m_{0}^{2}-\Sigma(m^{2}_{r})+i\epsilon} =
\frac{1}{2}\int \frac{d^{D}q}{(2 \pi)^{D}} \;
\frac{i}{q^{2}-m^{2}_{r}+i\epsilon}\, , \tseleq{b45}
\end{equation}
\noindent hence we see that $\Sigma$ is external--momentum
independent. If we had started with the renormalization
prescription: $i\Gamma^{(2)}_{r}(p^{2}=0)= - m^{2}_{r}$, we would
arrived at (\tseref{c1}) as well (this is not the case for
$N=1$!).

\vspace{3mm}

At finite temperature the strategy is analogous. Due to a doubling
of degrees of freedom, the full propagator is a $2\times 2$
matrix. The latter satisfies, similarly as at $T=0$, Dyson's
equation
\begin{equation}
\D = \D_{F} + \D_{F} \left(-i{\vect{\Sigma}} \right) \D\, .
\tseleq{bc5}
\end{equation}
\noindent An important point is that there exists a real,
non--singular matrix $\M $ (Bogoliubov matrix)\tsecite{LW,LB,TA}
having a property that
\begin{equation}
\D_{F} = \M \left( \begin{array}{cc}
                                   i\Delta_{F} & 0 \\
                                        0      & -i\Delta^{*}_{F}
                                    \end{array} \right) \M
\;\; \; \; \mbox{and} \; \; \; \;
{\vect{\Sigma}} = \M^{-1} \left( \begin{array}{cc}
                                   \Sigma_{T} & 0 \\
                                        0      & -\Sigma^{*}_{T}
                                    \end{array} \right) \M^{-1}\, .
\tseleq{m8}
\end{equation}
\noindent Here $\Delta_{F}$ is the standard Feynman propagator and
* denotes the complex conjugation. Consequently, the full matrix
propagator may be written as
\begin{equation}
\D = \M \left( \begin{array}{cc}
              \frac{i}{p^{2}-m_{0}^{2}-\Sigma_{T}+i\epsilon} & 0 \\
                             0 &
\frac{-i}{p^{2}-m_{0}^{2}-\Sigma^{*}_{T}-i\epsilon}
                  \end{array} \right) \M\, .
\tseleq{m9}
\end{equation}
\noindent Similarly as in many body systems, the position of the
(real) pole of $\D$ in $p^{2}$ fixes the temperature--dependent
effective mass $m_{r}(T)$\tsecite{LB,FW}. The latter is determined
by the equation
\begin{equation}
m_{r}^{2}(T) = m_{0}^{2} +
\mbox{Re}\left(\Sigma_{T}(m^{2}_{r}(T))\right)\, . \tseleq{m10}
\end{equation}
\noindent From the explicit form of $\M$ it is possible to
show\tsecite{LW,LB} that
$\mbox{Re}{\vect{\Sigma}}_{11}=\mbox{Re}\Sigma_{T}$. As before,
the structure of the proper self--energy can be deduced from the
corresponding Dyson--Schwinger equation. Following the usual
real--time formalism convention (type--1 vertex $\sim
-i\lambda_{0}$, type--2 vertex $\sim i\lambda_{0}$ ), the former
reads:

\begin{figure}[h]
\begin{center}
\leavevmode
\hbox{%
\epsfxsize=12.4cm \epsffile{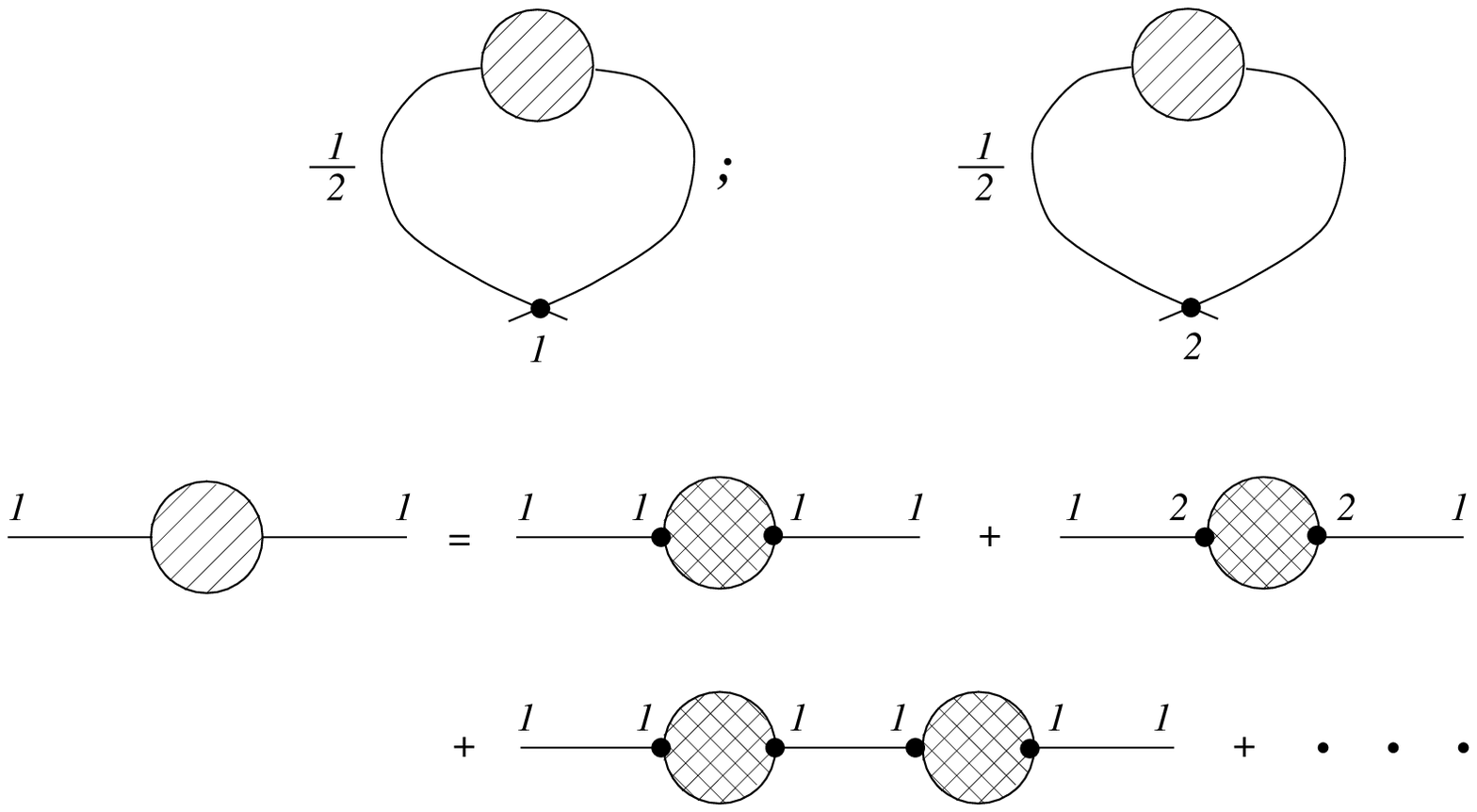}} \label{fig6}
\end{center}
\setlength{\unitlength}{1mm}
\begin{picture}(10,7)
\put(21,65){${-i\vect{\Sigma}}_{11}=$}
\put(76,65){${-i\vect{\Sigma}}_{22}=$} \put(0,47.5){where}
\put(141.5,64){(4.8)}
\end{picture}
\end{figure}
\addtocounter{equation}{1}
%
%
\noindent and similarly for $\D_{22}$. In (4.8) we have omitted
diagrams which are of order ${\cali{O}}(1/N)$ or less. Note that
the fact that no setting sun diagrams are present implies that the
off--diagonal elements of ${\vect{\Sigma}}$ are zero. Inspection
of Eq.(4.8) reveals that
\begin{equation}
{\vect{\Sigma}}_{11}= \frac{\lambda_{0}}{2}\; \int
\frac{d^{D}q}{(2\pi)^{D}} \; {\D}_{11}(q;T) \; \; \; \mbox{and} \;
\; \; {\vect{\Sigma}}_{22}= -\frac{\lambda_{0}}{2}\; \int
\frac{d^{D}q}{(2\pi)^{D}} \; {\D}_{22}(q;T)\, . \tseleq{m11}
\end{equation}
\noindent It directly follows from Eq.(\tseref{m11}) that both
${\vect{\Sigma}}_{11}$ and ${\vect{\Sigma}}_{22}$ are
external--momentum independent and real \footnote{Reality of
${\vect{\Sigma}}_{11}$ can be most easily seen from the
largest--time equation\tsecite{PJ}. The LTE states that
${\vect{\Sigma}}_{11} + {\vect{\Sigma}}_{22} +
{\vect{\Sigma}}_{12} + {\vect{\Sigma}}_{21} = 0$.  Because no
setting sun graphs are present, ${\vect{\Sigma}}_{12} +
{\vect{\Sigma}}_{21} = 0$, on the other hand ${\vect{\Sigma}}_{11}
+ {\vect{\Sigma}}_{22} = 2i\mbox{Im}{\vect{\Sigma}}_{11}$ (see
(\tseref{m8})).}. If we define $\Sigma_{T}(m_{r}^{2}(T)) =
\lambda_{0}\;  {\cali{M}}_{T}(m_{r}^{2}(T))$, then
Eq.(\tseref{m10}) through Eq.(\tseref{m11}) implies that
\begin{equation}
m_{r}^{2}(T) = m^{2}_{0} +  \lambda_{0}\;
{\cali{M}}_{T}(m_{r}^{2}(T))\, . \tseleq{m12}
\end{equation}
\noindent A resumed version of $\D_{11}$ is easily obtainable from
(\tseref{m9})\tsecite{LW,LB} and consequently (\tseref{m11})
yields
\begin{eqnarray}
{\cali{M}}_{T}(m_{r}^{2}(T)) &=& \frac{1}{2}\; \int
\frac{d^{D}q}{(2\pi)^{D}}\; \left\{ \frac{i}{q^{2} - m^{2}_{r}(T) +
i\epsilon} \; \; +\; (4\pi)\; \delta^{+}(q^{2}-m^{2}_{r}(T))\;
\frac{1}{e^{q_{0}\beta}-1} \right\}\nonumber \\
&=&- \int \frac{d^{D}q}{(2\pi)^{D}}\;
\frac{\varepsilon
(q_{0})}{e^{q_{0}\beta}-1}\; \mbox{Im}\frac{1}{q^{2}-m_{r}^{2}(T)
+i\epsilon}. \tselea{m14}
\end{eqnarray}
\noindent Let us remark that (\tseref{m14}) is manifestly
independent of any particular real--time formalism version.

\vspace{3mm}

In passing it may be mentioned that because
${\vect{\Sigma}}_{11}(m^{2}_{r})$ is momentum independent, the
wave function renormalization $Z_{\phi} = 1$. (The
K{\"a}llen--Lehmann representation requires the renormalized
propagator to have a pole of residue $i$ at $p^{2}=m^{2}_{r}$. The
former in turn implies that
$Z_{\phi}=(1-{\vect{\Sigma}}_{11}'(p^{2})|_{p^{2}=m^{2}_{r}})^{-1}
=1$.) Trivial consequence of the foregoing fact is that
$\Gamma^{(2)}_{r}=\Gamma^{(2)}$ and
$\Gamma^{(4)}_{r}=\Gamma^{(4)}$.

\subsection{Coupling constant renormalization} \label{PE32}

Let us choose the coupling constant to be defined at $T=0$. This
will have the advantage that the high temperature expansion of the
pressure (see Section \ref{HTE}) will become more transparent. In
addition, such a choice will allow us to select safely the part of
the parameter space in which spontaneous symmetry breakdown is not
possible.
%
An alternative renormalization
procedure based on the affective action is presented
in\tsecite{HS}.

\vspace{3mm}

By assumption the fields $\phi_{a}$ have non--vanishing masses, so
we can safely choose the renormalization prescription for
$\lambda_{r}$ at $s = 0$ ($s$ is the standard Mandelstam
variable). For example, one may require that for the scattering
$aa \rightarrow bb$
\begin{equation}
\Gamma^{(4)}(s=0)= -\lambda_{r}/N, \; \; \; \; \; (b \not= a)\, .
 \tseleq{m13}
\end{equation}
\noindent The formula (\tseref{m13}) clearly agrees with the tree
level value $\Gamma_{tree}^{(4)\;aabb}(s=0)= -\lambda_{0}/N$. Let
us also mention   that Ward's  identities corresponding  to the
internal $O(N)$  symmetry enforce      $\Gamma^{(4)\;  aaaa}$ to
obey  the constraint\footnote{Actually,    Ward's identities
read\tsecite{PC} $\int        d^{D}x          \;
\frac{\delta\Gamma[\phi]}{\delta \phi_{a}(x)}\;\phi_{b}(x)       =
\int      d^{D}x        \; \frac{\delta\Gamma[\phi]}{\delta
\phi_{b}(x)}\;\phi_{a}(x)$   (here $\phi_{a} =  \frac{\delta
W}{\delta J_{a}}$; $W$  is  the generating functional  of
connected  Green's functions).  Performing successive variations
with respect to $\phi_{a}(v), \phi_{a}(z), \phi_{a}(y)$ and
$\phi_{b}(w)$, taking the  Fourier transform, and setting the
physical condition $\phi_{c} = 0$, we get directly
(\tseref{m134}).}
\begin{eqnarray}
\Gamma^{(4)\; aaaa}(p_{1}; p_{2}; p_{3}; p_{4}) &=& \Gamma^{(4)\;
bbaa}(p_{1}; p_{2}; p_{3}; p_{4}) + \Gamma^{(4)\; baba}(p_{1};
p_{2};
p_{3}; p_{4})\nonumber \\
 &+& \Gamma^{(4)\; baab}(p_{1}; p_{2}; p_{3};
p_{4})\, , \tselea{m134}
\end{eqnarray}
\noindent for any  $b \not=  a$.  The structure of $\Gamma^{(4)}$
is encoded in the Dyson--Schwinger equation (4.14) (see also
\tsecite{IZ,PC}).
\begin{figure}[h]
\begin{center}
\leavevmode
\hbox{%
\epsfxsize=14 cm \epsffile{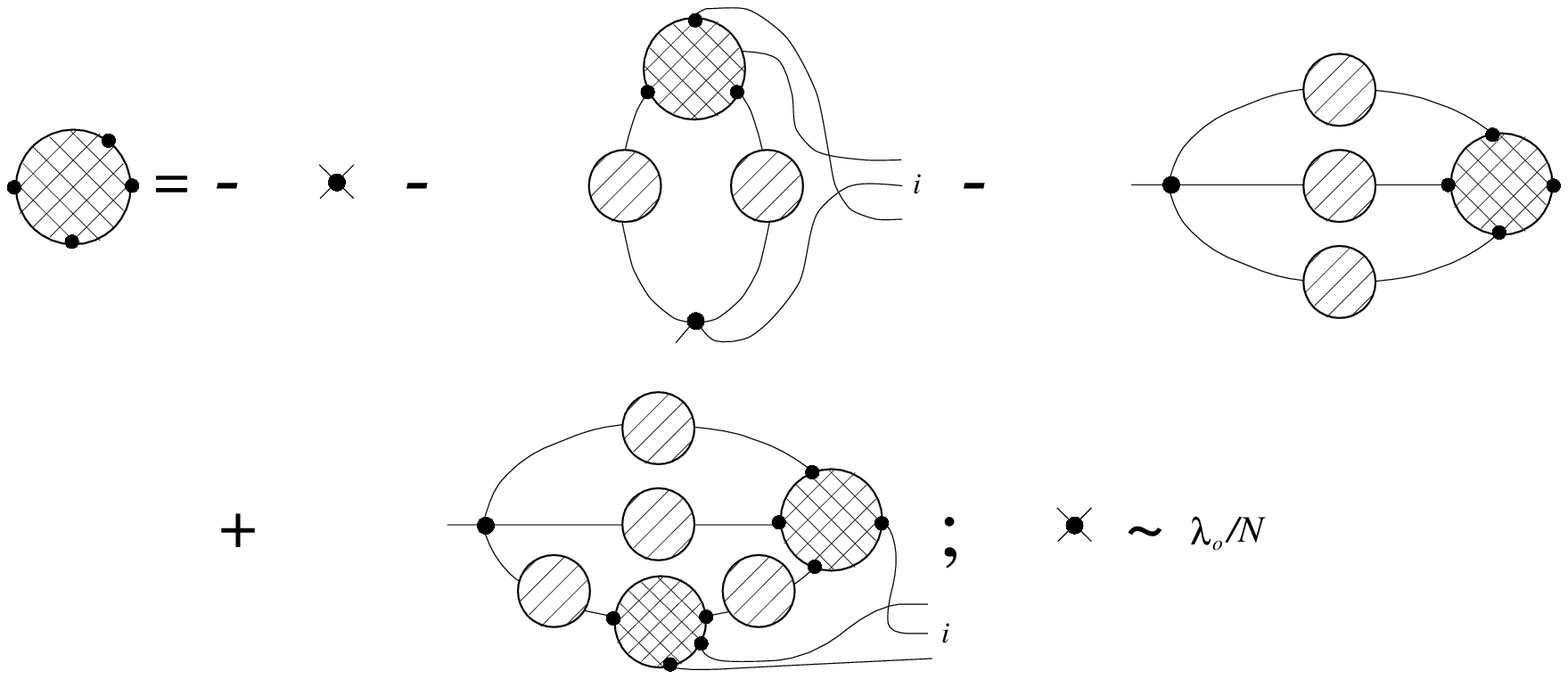}}
\end{center}
\setlength{\unitlength}{1mm}
\begin{picture}(20,7)
\put(45.5,54.5){$\frac{i}{2}\sum_{i=1}^{3}$}
\put(102.5,54.5){$\frac{1}{6}$}
\put(33.5,23.5){$\frac{1}{2}\sum_{i=1}^{3}$}
\put(138,23){(4.14)}
\end{picture}
\end{figure}
\addtocounter{equation}{1}
In the latter the sum $\sum_{i=1}^{3}$ schematically represents a
summation over $s$, $t$ and $u$ scattering channels. For clarity's
sake the internal indices are suppresses. Similarly as before, we
can argue that both the third and fourth graphs contribute at most
$N^{-2}$, whilst the second (``fish") graph may contribute up to
order $N^{-1}$. So in the large--$N$ limit the last three diagrams
may be neglected, provided we keep in the $4$--point vertex
function only graphs proportional to $N^{-1}$. However, the former
can be only fulfilled if we retain such a ``fish" graph where
summation over internal index on the loop is allowed. Remaining
two graphs in the sum $\sum_{i=1}^3$ (i.e.,  $t$ and $u$
scattering channels) are suppressed by the factor $N^{-1}$ as the
internal index on the loop is fixed. In this way we are left with
the relation
\begin{eqnarray}
&&\Gamma^{(4)\; aabb}(s=0)\nonumber \\
&&\mbox{\hspace{0 mm}} = -\frac{\lambda_{0}}{N} - \left.
\frac{i\lambda_{0}}{2N} \sum_{c \not= b}\int
\frac{d^{D}q}{(2\pi)^{D}}\; \Gamma^{(4)\; bbcc}(s) \;
\frac{i}{\left( q^{2} - m_{r}^{2} + i\epsilon \right)}\;
\frac{i}{\left( (q -Q)^{2} - m_{r}^{2} + i\epsilon
\right)}\right|_{s=0}
\nonumber \\
&&\mbox{\hspace{0 mm}} =- \frac{\lambda_{0}}{N} -
\frac{\lambda_{0} \lambda_{r}(N-1)}{2N^{2}} \int^{1}_{0}dx \int
\frac{d^{D}q}{(2\pi)^{D}}\;\left. \frac{i}{\left( q^{2} -
m_{r}^{2} + x(1-x)s + i\epsilon \right)^{2}} \right|_{s=0}\, ,\nonumber\\
\tseleq{m145}
\end{eqnarray}
\noindent with $Q=p_{1}+p_{2}$ and $s=Q^{2}$, $p_{1}, p_{2}$ are
the external momenta. To leading order in $1/N$ we may
equivalently write
\begin{equation}
\lambda_{r} = \lambda_{0} + \lambda_{0}\lambda_{r}\;
{\cali{M}}'(m^{2}_{r})\, , \tseleq{m146}
\end{equation}
\noindent the prime means differentiation with respect to
$m_{r}^{2}$; ${\cali{M}}(m^{2}_{r})$ is defined by (\tseref{b45}).
Evaluating explicitly ${\cali{M}}'(m^{2}_{r})$, we get from
(\tseref{m146})
\begin{equation}
\lambda_{0} =
\frac{\lambda_{r}}{1-\lambda_{r}\Gamma(2-\frac{D}{2})\;(m_{r})^{D-4}/2\;(4
\pi)^{\frac{D}{2}}}\, . \tseleq{m1465}
\end{equation}
\noindent Assuming that both $\lambda_{0} \geq 0$ and $\lambda_{r}
\geq 0$ (note $\lambda_r <0$ would be incompatible with
$\lambda_{0} \geq 0$ and $m_r^2 >0$), we can infer from
(\tseref{m1465}) that
\begin{equation}
0 \leq \lambda_{r} \leq
\frac{2(4\pi)^{\frac{D}{2}}(m_{r})^{4-D}}{\Gamma(2-\frac{D}{2})}\,
,
\end{equation}
\noindent and so for $D=4$ we inevitably get that $\lambda_{r}=
0$. The latter indicates that the theory is
trivial\tsecite{ID1,PR,BM}, or, in other words, the $O(N)\;
\phi^{4}$ theory is a renormalized free theory in the large--$N$
limit. This conclusion is also consistent with the observation
that the theory does not posses any non--trivial UV fixed point in
the large--$N$ limit\tsecite{PR,BM,F2,FSW}.

\vspace{3mm}

On the other hand, if we were assuming that $\lambda_{0} < 0$, we
would get indeed a non--trivial renormalized field theory in $D=4$
(actually, from (\tseref{m1465}) we see that $\lambda_{0}
\rightarrow 0_{-}$ , provided that $\lambda_{r}$ is fixed and
positive and $D \rightarrow 4_{-}$).  However, as it was pointed
out in refs.\tsecite{ID1,PR,BM,AKS}, such a theory is
intrinsically unstable as the ground--state energy is unbounded
from below. This is reflected, for instance, in the existence of
tachyons in the theory\tsecite{ID1,BM,AKS,RGR}, therefore the case
with negative $\lambda_{0}$ is clearly inconsistent.

\vspace{3mm}

The straightforward remedy for this situation was suggested by
Bardeen and Moshe\tsecite{BM}. They showed that the only
meaningful (stable) $O(N)\; \phi^{4}$ theory in the large--$N$
limit is that with $\lambda_{r}, \lambda_{0} \geq 0$. This is
provided that we view it as an effective field theory at momenta
scale small compared to a fixed UV cut--off $\Lambda$. The
cut--off itself is further determined by (\tseref{m146}) because
in that case (assuming $m_{r} \ll \Lambda$)
\begin{equation}
\lambda_{0} = \frac{\lambda_{r}}{1- \frac{\lambda_{r}}{32
\pi^{2}}\mbox{ln}(\frac{\Lambda^{2}}{m^{2}_{r}})}\, ,
\label{cutoff}
\end{equation}
\noindent which implies that for $\lambda_{r}, \lambda_{0} \geq 0$
we have $\Lambda^{2} < m^{2}_{r}\;\mbox{exp}(\frac{32
\pi^{2}}{\lambda_{r}})$. The case $\Lambda^{2} =
m^{2}_{r}\;\mbox{exp}(\frac{32 \pi^{2}}{\lambda_{r}})$ corresponds
to the Landau ghost\tsecite{AC2} (tachyon pole\tsecite{ID1,BM}).
For reasonably small $\lambda_{r}$, $\Lambda$ is truly
huge\footnote{For example, if $\lambda_{r}=1$ and $m_{r} \approx
100$MeV, we get $\Lambda < 10^{141}$MeV or equivalently $\Lambda <
10^{131}$K (this is far beyond the Planck temperature -
$10^{32}$K).} and so it does not represent any significant
restriction.

\vspace{3mm}

In passing we may observe that from (\tseref{c1}) and
(\tseref{cutoff})
\begin{equation}
\frac{m_r^2}{\lambda_r} = \frac{m_0^2}{\lambda_0} + \frac{1}{23
\pi^2} \ \Lambda^2 \, , \label{invariant}
\end{equation}
\noindent  and so the fraction $m_r^2/\lambda_r$ is
renormalization invariant. It was argued in\tsecite{BM} that for
the part of the parameter space where $\lambda_0 > 0$ and
$m_r^2/\lambda_r > 0$ the ground state is $O(N)$ symmetric,
Goldstone phenomena cannot materialize and hence the expectation
value of the field is zero. The latter fact has been implicitly
used, for instance, in the derivation our Dyson--Schwinger
equations. To avoid a delicate discussion of the phase structure
of the $O(N) \ \phi^4$ theory and to emphasize our primary
objective - i.e. hydrostatic pressure calculation, we confine
ourselves to the parameter space defined above. Such an effective
theory will provide a suitable playground to explore all the basic
salient points involved in the hydrostatic pressure calculation.
Furthermore, because the mass--shift equation (gap equation) has a
particularly simple form in this case the hight--$T$ analysis of
the hydrostatic pressure will be easy to perform.

%
%

\subsection{Resumed hydrostatic pressure} \label{PE33}

\begin{figure}[h]
\vspace{4mm} \epsfxsize=11cm \centerline{\epsffile{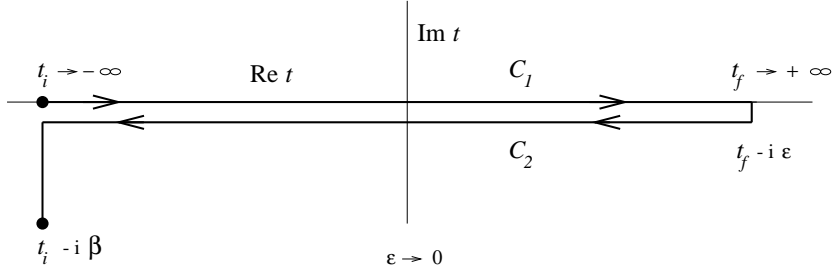}}
\caption{\em The Keldysh--Schwinger time path.} \label{fig18}
\vspace{4mm}
\end{figure}

\begin{figure}[h]
\begin{center}
\leavevmode
\hbox{%
\epsfxsize=11cm \epsffile{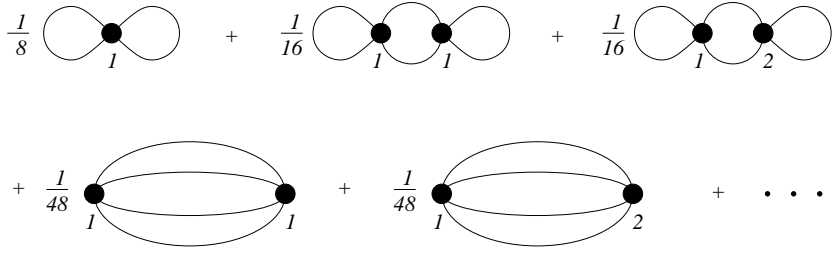}} \caption{\em First few bubble
diagrams in the $\Phi$ expansion.} \label{fig3}
\end{center}
\end{figure}

The partition function $Z$ has a well known path-integral
representation at finite temperature, namely
\begin{eqnarray}
&&Z[T] = \mbox{exp}(\Phi[T]) = \int{\cali{D}}\phi\;
\mbox{exp}(iS[\phi;T])\, ,\nonumber \\ &&S[\phi;T] =
\int_{C}d^{D}x\; {\cali{L}}(x)\, . \tselea{b1}
\end{eqnarray}
\noindent Here $\Phi= -\beta \Omega$ is a Massieu function (the
Legendre transform of the entropy)\tsecite{LW,GM,Cub,Call1} and
$\int_{C}d^{D}x = \int_{C}dx_{0}\int_{V}d^{D-1}{\vect{x}}$ with
the subscript $C$ suggesting that the time runs along some contour
in the complex plane. In the real--time formalism, which we adopt
throughout, the most natural version is the so called
Keldysh--Schwinger one\tsecite{LW,LB}, which is represented by the
contour in Fig.\ref{fig18}.
\noindent Let us mention that the fields within the path--integral
(\tseref{b1}) are further restricted by the periodic boundary
condition (KMS condition)\tsecite{LW,LB,TA} which in our case
reads:
\begin{displaymath}
\phi_{a}(t_{i}-i\beta, {\vect{x}})= \phi_{a}(t_{i}, {\vect{x}})\,
.
\end{displaymath}
\noindent As explained in Section \ref{PE2}, we can use for a
pressure calculation the canonical energy--momentum tensor
$\Theta^{\mu \nu}_{c}$. Employing for $\Theta^{\mu \nu}_{c}(x)$
its explicit form (\tseref{7}) together with (\tseref{11}), one
may write
\begin{equation}
\langle \Theta^{\mu \nu}_{c} \rangle = \frac{N}{2}\; \int
\frac{d^{D}q}{(2\pi)^{D}}(2q^{\mu}q^{\nu}-g^{\mu
\nu}(q^{2}-m^{2}_{0}))\; \D_{11}(q;T) \; \; +
\frac{\lambda_{0}}{8N} g^{\mu \nu} \left\langle \left(
\sum_{a=1}^{N}\phi_{a}^{2}(0)\right)^{2} \right\rangle \, ,
\tseleq{b2}
\end{equation}
\noindent where $\D_{11}$ is the Dyson--resumed thermal
propagator\tsecite{LW,LB}, i.e.
\begin{equation}
\D_{11}(q;T) = \frac{i}{q^{2}-m^{2}_{r}(T)+i\epsilon} \; \; +
(2\pi) \; \delta (q^{2}-m^{2}_{r}(T))\;
\frac{1}{e^{|q_{0}|\beta}-1}\, . \tseleq{b3}
\end{equation}
\noindent Note that we have exploited in (\tseref{b2}) the fact
that the expectation value of $\Theta^{\mu \nu}_{c}(x)$ is $x$
independent. On the other hand, in (\tseref{b3}) we have used the
fact that $m^{2}_{r}$ is $q$ independent. In order to calculate
the expectation value of the quartic term in Eq.(\tseref{b2}), let
us observe (c.f. (\tseref{b1})) that the derivative of $\Phi$ with
respect to the bare coupling $\lambda_{0}$ (taken at fixed
$m_{0}$) gives
\begin{equation}
\frac{\partial\Phi[T]}{\partial \lambda_{0}}= - \frac{i}{8\; N}\;
\int_{C} d^{D} x \left\langle \left(\sum_{a=1}^{N}\phi_{a}^{2}(0)
\right)^{2} \right\rangle \nonumber \nonumber \, ,
\end{equation}
\noindent which implies that
\begin{equation}
\left\langle \left(\sum_{a=1}^{N}\phi_{a}^{2}(0) \right)^{2}
\right\rangle = - \frac{N8}{\beta
V}\;\frac{\partial\Phi[T]}{\partial \lambda_{0}}\, .
\tselea{b4}
\end{equation}
\noindent The key point now is that we can calculate $\Phi[T]$ in
a non--perturbative form. (The latter is based on the fact that we
know the Dyson--resumed propagator $\D_{11}(q;T)$ (see
(\tseref{b3}).) Indeed, taking derivative of $\Phi$ with respect
to $m_{0}^{2}$ (keeping $\lambda_{0}$ fixed) we obtain
\begin{eqnarray}
\frac{\partial \Phi [T]}{\partial m_{0}^{2}}&=&
-\frac{iN}{2}\int_{C}d^{D}x \left\langle \phi^{2}(0) \right\rangle
= -\frac{\beta V N}{2} \int
\frac{d^{D}q}{(2\pi)^{D}}\; \D_{11}(q;T)\nonumber \\
&=& - \beta V N\; {\cali{M}}_{T}(m_{r}^{2}(T))\, , \tselea{b5}
\end{eqnarray}
\noindent thus
\begin{equation}
\Phi[T;\lambda_{0}; m_{0}^{2}] = \beta V N \;
\int_{m_{0}^{2}}^{\infty}
d{\hat{m}}_{0}^{2}\;{\cali{M}}_{T}({\hat{m}}_{r}^{2}(T))\; \; +
\Phi[T;\lambda_{0}; \infty]\, .
\tseleq{b6}
\end{equation}
\noindent Let us note that $\Phi[T;\lambda_{0}; \infty]$ is
actually zero\footnote{To be precise, we should also include in
Fig.\ref{fig3} an (infinite) circle diagram corresponding to the
free pressure\tsecite{LW,EM}. However, the later is $\lambda_{0}$
independent (although $m_{0}$ dependent) and so it is irrelevant
for the successive discussion (c.f. Eq.(\tseref{b4})).} because
$\Phi[T;\lambda_{0};m^{2}_{0}]$ has the standard loop
expansion\tsecite{LW,PC} depicted in Fig.4.
It is worth mentioning that in the latter expansion one must
always have at least one type--1 vertex\tsecite{LW}. The RHS of
Fig.\ref{fig3} clearly tends to zero for $m_{0} \rightarrow
\infty$ as all the (free) thermal propagators from which the
individual diagrams are constructed tend to zero in this limit.
The former result can be also deduced from the CJT effective
action formalism\tsecite{CJT} or from a heuristic argumentation
based on a thermodynamic pressure\tsecite{ID1}. Note that in the
large--$N$ limit the fourth and fifth diagrams in Fig.\ref{fig3}
must be omitted.

\vspace{3mm}

The expectation value (\tseref{b4}) can be now explicitly written
as
\begin{equation}
\left\langle \left( \sum_{a=1}^{N}\phi_{a}^{2}(0) \right)^{2}
\right\rangle = 8N^{2}\; \int_{m_{0}^{2}}^{\infty}
d{\hat{m}}_{0}^{2}\; \int \frac{d^{D}q}{(2\pi)^{D}}\;
\frac{\varepsilon(q_{0})}{e^{q_{0}\beta}-1}\; \mbox{Im}
\left(\frac{\frac{\partial
\Sigma_{T}({\hat{m}}_{r}^{2}(T))}{\partial \lambda_{0}}}{(q^{2} -
{\hat{m}}^{2}_{r} +i\epsilon)^{2}}\right)\, .
\tseleq{b65}
\end{equation}
\noindent In fact, the differentiation of the proper self--energy
in (\tseref{b65}) can be carried out easily. Using (\tseref{m12}),
we get
\begin{displaymath}
\frac{\partial \Sigma_{T}}{\partial \lambda_{0}} =
\frac{\Sigma_{T}}{\lambda_{0}} + \lambda_{0}{\cali{M}}_{T}' \;
\frac{\partial \Sigma_{T}}{\partial \lambda_{0}} \;\; \Rightarrow
\;\; \frac{\partial \Sigma_{T}}{\partial \lambda_{0}} =
\frac{\Sigma_{T}}{\lambda_{0}(1-\lambda_{0}{\cali{M}}'_{T})}\, .
\end{displaymath}
\noindent From Eq.(\tseref{m12}) it directly follows that
\begin{displaymath}
\frac{dm_{r}^{2}(T)}{dm_{0}^{2}} = \frac{1}{(1-
\lambda_{0}\;{\cali{M}}'_{T})}\, ,
\end{displaymath}
\noindent which, together with the definition of ${\cali{M}}_{T}$, gives
\begin{eqnarray}
\left\langle \left( \sum_{a=1}^{N} \phi_{a}^{2}(0) \right)^{2}
\right\rangle &=& 8N^{2}\; \int_{m_{r}^{2}(T)}^{\infty}
d{\hat{m}}_{r}^{2}\; \int \frac{d^{D}q}{(2 \pi)^{D}} \;
\frac{\varepsilon(q_{0})}{e^{q_{0}\beta} -1} \;
\mbox{Im}\frac{{\cali{M}}_{T}({\hat{m}}_{r}^{2})}{(q^{2}-
{\hat{m}}_{r}^{2} +i\epsilon)^{2}}\nonumber \\
&=& -\; 8N^{2} \; \int_{m_{r}^{2}(T)}^{\infty}
d{\hat{m}}_{r}^{2}\; {\cali{M}}_{T}({\hat{m}}_{r}^{2}) \;
\frac{\partial {\cali{M}}_{T}({\hat{m}}_{r}^{2})}{\partial
{\hat{m}}_{r}^{2}}\nonumber \\
&=& 4N^{2}\; {\cali{M}}_{T}^{2}(m_{r}^{2}(T))\, ,
\tselea{b8}
\end{eqnarray}
\noindent where we have exploited in the last line the fact that
${\cali{M}}_{T}^{2}(m_{r}^{2} \rightarrow \infty)=0$. Let us mention that
the crucial point in the previous manipulations was that $m_{r}$ is both
real and momentum independent. Collecting our results together, we can
write for the hydrostatic pressure per particle (cf. Eq(\tseref{ppp9}))
\begin{eqnarray}
{\cali{P}}(T)-{\cali{P}}(0) &=& -\; \frac{1}{(D-1)N}\; \left(
\langle \Theta^{i}_{c \;i} \rangle - \langle 0| \Theta^{i}_{c\; i}
| 0 \rangle
\right)\nonumber \\
&=& +\; \frac{1}{2}\; \int \frac{d^{D}q}{(2 \pi)^{D-1}}\;
\left( \frac{2 {\vect{q}}^{2}}{(D-1)}  \right)\;
\frac{\varepsilon(q_{0})}{e^{q_{0}\beta}-1}\;
\delta(q^{2}-m_{r}^{2}(T))\nonumber \\
&& -\; \frac{1}{2}\; \int \frac{d^{D}q}{(2
\pi)^{D-1}}\;
\left( \frac{2 {\vect{q}}^{2}}{(D-1)} \right)\;
\delta^{+}(q^{2}-m_{r}^{2}(0))\nonumber \\
&&+ \frac{1}{2\lambda_{0}}\left( \Sigma^{2}_{T}(m_{r}^{2}(T)) -
\Sigma^{2}(m_{r}^{2}(0)) \right)\, . \tselea{b9}
\end{eqnarray}
\noindent Applying Green's theorem to the last two integrals and
eliminating the surface terms (for details see Appendix A) we find
\begin{eqnarray} {\cali{P}}(T) - {\cali{P}}(0) &=& \frac{1}{2}\; \int
\frac{d^{D}q}{(2 \pi)^{D-1}}\; \frac{\varepsilon(q_{0})}{e^{q_{0}\beta}-1}\;
\theta(q^{2}-m^{2}_{r}(T))\nonumber \\ &&- \frac{1}{2} \; \int
\frac{d^{D}q}{(2 \pi)^{D-1}}\; \theta(q_{0}) \;
\theta(q^{2}-m^{2}_{r}(0))\nonumber \\
&&+\frac{1}{2\lambda_{0}}\left( \Sigma^{2}_{T}(m_{r}^{2}(T))-
\Sigma^{2}(m_{r}^{2}(0))\right) \nonumber \\ &&\nonumber \\
&&\mbox{\hspace{-2.5cm}} = \ {\cali{N}}_{T}(m_{r}^{2}(T)) -
{\cali{N}}(m_{r}^{2}(0)) + \frac{1}{2\lambda_{0}} \left(
\Sigma^{2}_{T}(m_{r}^{2}(T))- \Sigma^{2}(m_{r}^{2}(0))\right)\, ,
\tseleq{b10}
\end{eqnarray}
\noindent where we have introduced new functions
${\cali{N}}_{T}(m^{2}_{r}(T)$ and ${\cali{N}}(m^{2}_{r})$;
\begin{eqnarray}
{\cali{N}}_{T}(m^{2}_{r}(T)) &=& \frac{1}{2}\; \int
\frac{d^{D}q}{(2 \pi)^{D-1}}\; \frac{\varepsilon(q_{0})}{e^{q_{0}\beta}-1}\;
\theta(q^{2}-m^{2}_{r}(T)) \nonumber \\
{\cali{N}}(m^{2}_{r}) &=& \lim_{T \rightarrow \;
0}{\cali{N}}_{T}(m^{2}_{r}(T))\, . \tselea{b101}
\end{eqnarray}
\noindent Eq.(\tseref{b10}) can be rephrased into a form which exhibits an
explicit independence of bar quantities. Using the trivial identity:
\begin{eqnarray}
&&\frac{1}{2\lambda_{0}}\; \left( \Sigma^{2}_{T}(m_{r}^{2}(T))-
\Sigma^{2}(m_{r}(0))\right)\nonumber \\
&& \mbox{\hspace{1.5cm}}= \frac{1}{2\lambda_{0}}\; \left(
\Sigma_{T}(m_{r}^{2}(T))-\Sigma(m^{2}_{r}(0))\right)\left(
\Sigma_{T}(m_{r}^{2}(T))+\Sigma(m^{2}_{r}(0))\right) \nonumber \\
&& \mbox{\hspace{1.5cm}}= \frac{\delta m^{2}(T)}{2}\;\left(
{\cali{M}}_{T}(m_{r}^{2}(T))+{\cali{M}}(m^{2}_{r}(0))\right)\, ,
\tselea{b111}
\end{eqnarray}
\noindent we get
\begin{eqnarray} &&\mbox{\hspace{-14mm}}{\cali{P}}(T) - {\cali{P}}(0)\nonumber \\
&&\mbox{\hspace{-6mm}} = \ {\cali{N}}_{T}(m_{r}^{2}(T)) -
{\cali{N}}(m_{r}^{2}(0))+ \frac{\delta m^{2}(T)}{2}\;\left(
{\cali{M}}_{T}(m_{r}^{2}(T))+{\cali{M}}(m^{2}_{r}(0))\right)\, ,
\tseleq{b11}
\end{eqnarray}
\noindent where $\delta m^{2}(T)= m_{r}^{2}(T)-m_{r}^{2}(0)$. Let
us finally mention that the finding (\ref{b9}) is an original
result of this paper. The result (\tseref{b11}) has been
previously obtained by authors\tsecite{ID1} in the purely
thermodynamic pressure framework.

\section{High--temperature
expansion of the hydrostatic pressure in $D=4$ }\label{HTE}

In order to obtain the high--temperature expansion of the pressure
in $D=4$, it is presumably the easiest to go back to equation
(\tseref{b9}) and employ identity (\tseref{b111}). Let us split
this task into two parts.  We firstly evaluate the integrals with
potentially UV divergent parts using the dimensional
regularization. The remaining integrals, with the Bose--Einstein
distribution insertion, are safe of UV singularities and can be
computed by means of the Mellin transform technique.

\vspace{3mm}

Inspecting (\tseref{b9})  and (\tseref{b111}), we observe that the
only UV divergent contributions come from the integrals:
\begin{eqnarray}
&+& \frac{1}{(D-1)} \int \frac{d^{D}q}{(2\pi)^{D-1}} \; {\vect{q}}^{2}\;
\left( \delta^{+}(q^{2}-m^{2}_{r}(T)) - \delta^{+}(q^{2}-m^{2}_{r}(0))
\right)\nonumber \\
&+& \frac{\delta m^{2}(T)}{4} \int \frac{d^{D}q}{(2\pi)^{D}} \;
\left( \frac{i}{q^{2} - m^{2}_{r}(T) + i\epsilon } +
\frac{i}{q^{2}-m^{2}_{r}(0)+i\epsilon}\right)\, ,
\tselea{4.1}
\end{eqnarray}
\noindent which, if integrated over, give
\begin{eqnarray}
\mbox{(\tseref{4.1})} = &+&
\frac{\Gamma(\frac{-D}{2})\Gamma(\frac{D}{2}+\frac{1}{2})}{(D-1)
\Gamma(\frac{D-1}{2}) (4 \pi)^{\frac{D}{2}}} \left(
(m^{2}_{r}(T))^{\frac{D}{2}} -
(m^{2}_{r}(0))^{\frac{D}{2}}\right)\nonumber \\
&+&\frac{\delta m^{2}(T)\; \Gamma(1-\frac{D}{2})}{4
(4\pi)^{\frac{D}{2}}} \left( (m^{2}_{r}(T))^{\frac{D}{2}-1} +
(m^{2}_{r}(0))^{\frac{D}{2}-1}\right)\, . \tselea{4.2}
\end{eqnarray}
\noindent Taking the limit $D=4-2\varepsilon \rightarrow 4$ and using
expansions
\begin{eqnarray*}
\Gamma(-n + \varepsilon) &=& \frac{(-1)^{n}}{n!} \left(\frac{1}{
\varepsilon} + \sum_{k=1}^{n} \frac{1}{k} - \gamma +
{\cali{O}}(\varepsilon) \right)\, ,\nonumber \\
a^{x+\varepsilon} &=&  a^{x} \left( 1 + \varepsilon \;\mbox{ln}a +
{\cali{O}}(\varepsilon^{2}) \right)\, ,
\end{eqnarray*}
\noindent ($\gamma$ is the Euler--Mascheroni constant) we are
finally left with
\begin{equation}
\left.\mbox{(\tseref{4.1})}\right|_{D \rightarrow 4} = -
\frac{m^{2}_{r}(0) m^{2}_{r}(T)}{64 \;\pi^{2}} \; \mbox{ln}\left(
\frac{m^{2}_{r}(T)}{m^{2}_{r}(0)}\right) + \delta
m^{2}(T)\;(m_{r}^{2}(T) + m^{2}_{r}(0))\; \frac{1}{128\;
\pi^{2}}\, . \tseleq{4.3}
\end{equation}
\noindent The fact that we get finite result should not be
surprising as entire analysis of Section \ref{PE2} was made to
show that ${\cali{P}}(T)-{\cali{P}}(0)$ defined via $\Theta^{\mu
\nu}_{c}$ is finite in $D=4$.

\vspace{3mm}

We may now concentrate on the remaining terms in (\tseref{b9}),
the latter read (we might, and we shall, from now on work in
$D=4$)
\begin{equation}
\frac{1}{3} \int \frac{d^{4}q}{(2 \pi)^{3}} \; {\vect{q}}^{2}\;
\frac{1}{e^{|q_{0}|\beta}-1}\; \delta(q^{2}-m_{r}^{2}(T)) +
\frac{\delta m^{2}(T)}{4} \int \frac{d^{4}q}{(2\pi)^{3}} \;
\frac{1}{e^{|q_{0}|\beta}-1}  \delta(q^{2}-m_{r}^{2}(T))\, .
\tseleq{4.4}
\end{equation}
\noindent Our following strategy is based on the observation that the
previous integrals have generic form:
\begin{eqnarray}
I_{2\nu}(m_{r}) &=& \int \frac{d^{4}q}{(2\pi)^{3}}\; {\vect{q}}^{2\nu}\;
\frac{1}{e^{|q_{0}|\beta}-1}\;
\delta(q^{2}-m^{2}_{r})\, ,\nonumber\\
&=& \frac{m_{r}^{2+2\nu}}{2\; \pi^{2}}\; \int_{1}^{\infty} dx
\;(x^{2}-1)^{\frac{1+2\nu}{2}}\; \frac{1}{e^{xy}-1}\, ,
\tseleq{c14}
\end{eqnarray}
\noindent with $\nu =0,1$ and $y = m_{r}\beta$.  Unfortunately,
the integral (\tseref{c14}) cannot be evaluated exactly, however,
its small $y$ (i.e. high--temperature) behavior can be
successfully analyzed by means of the Mellin transform
technique\tsecite{LW,EM}. Before going further, let us briefly
outline the basic steps needed for such a small $y$ expansion.

\vspace{3mm}

The Mellin transform $\hat{f}(s)$ is done by the
prescription\tsecite{LW,EM,Br,WH,B,BB}:
\begin{equation}
\hat{f}(s)= \int_{0}^{\infty} dx\; x^{s-1}\; f(x)\, , \tseleq{c12}
\end{equation}
\noindent with $s$ being a complex number. One can easily check that the
inverse Mellin transform reads
\begin{equation}
f(x)= \frac{1}{i(2\pi)} \int_{-i\infty + a}^{i\infty +a} ds\;
x^{-s} \; \hat{f}(s)\, , \tseleq{c13}
\end{equation}
\noindent where the real constant $a$ is chosen in such a way that
$\hat{f}(s)$ is convergent in the neighborhood of a straight line
($-i\infty +a,\; i\infty +a$). So particularly if $f(x) =
\frac{1}{e^{xy}-1}$ one can find (\tsecite{B}; formula I.3.19)
that
\begin{equation}
\hat{f}(s) = \Gamma(s)\zeta(s) y^{-s}\, , \mbox{\hspace{1cm}}
(\mbox{Re}s > 1)\, , \tseleq{4.6}
\end{equation}
\noindent where $\zeta$ is the Riemann zeta function ($\zeta(s)=
\sum_{n=1}^{\infty}n^{-s}$). Now we insert the Mellin transform of $f(x) =
\frac{1}{e^{xy}-1}$ to (\tseref{c14}) and interchange integrals (this is
legitimate only if the integrals are convergent before the
interchange). As a result we have
\begin{equation}
\int_{0}^{\infty} dx \; g(x)\ \frac{1}{e^{xy}-1} = \int_{-i\infty
+a}^{i\infty + a} \frac{ds}{i(2\pi)} \; \Gamma(s)\zeta(s) y^{-s}
\hat{g}(1-s)\, , \tseleq{4.7}
\end{equation}
\noindent with $g(x) = \theta(x-1)\;(x^{2}-1)^{\frac{1+2\nu}{2}}$. Using
the tabulated result (\tsecite{BB}; formula 6.2.32) we find
\begin{equation}
\hat{g}(1-s) = \frac{1}{2} B(-\nu -1 + \mbox{$\frac{1}{2}s$};
\mbox{$\frac{3}{2}$} + \nu)\, , \mbox{\hspace{1.5cm}} (\mbox{Re}s
>2+2\nu)\, , \tseleq{4.8}
\end{equation}
\noindent with $B(\;;\;)$ being the beta function. Because the integrand
on the RHS of (\tseref{4.7}) is analytic for $\mbox{Re}s > 2 + 2\nu$ and
the LHS is finite, we must choose such $a$ that the integration is
defined. The foregoing is achieved choosing $\mbox{a} > 2+2\nu$. Another
useful expressions for $\hat{g}(1-s)$ are (\tsecite{BB}; formula I.2.34 or
I.2.37)
\begin{eqnarray*}
\hat{g}(1-s) &=& B(\mbox{$\frac{3}{2}$}+\nu; -2 - 2\nu +s) \;
{~}_{2}F_{1}[-\mbox{$\frac{1}{2}$} -\nu ; -2 - 2\nu +s;
-\mbox{$\frac{1}{2}$} -\nu +s; -1] \nonumber \\
&=& 2^{\frac{1}{2} +\nu} \; B(\mbox{$\frac{3}{2}$}+\nu; -2 - 2\nu
+s)\; {~}_{2}F_{1}[-\mbox{$\frac{1}{2}$} -\nu ;
\mbox{$\frac{3}{2}$} +\nu; -\mbox{$\frac{1}{2}$} -\nu +s;
\mbox{$\frac{1}{2}$}]\, ,
\end{eqnarray*}
\noindent where ${~}_{2}F_{1}$ is the (Gauss) hypergeometric
function\tsecite{BB}.  Using identity
\begin{displaymath}
\Gamma(2x) = \frac{2^{2x-1}}{\sqrt{\pi}}\;
\Gamma(x)\Gamma(x+\mbox{$\frac{1}{2}$})\, ,
\end{displaymath}
\noindent we can write
\begin{equation}
\mbox{(\tseref{4.7})} = \frac{\Gamma(\mbox{$\frac{3}{2}$} +\nu)}{4
\sqrt{\pi}}\; \int_{-i\infty + a}^{i\infty +a} \frac{ds}{i(2\pi)}
\Gamma(\mbox{$\frac{1}{2}$}s) \zeta(s) \left(\mbox{$\frac{1}{2}$}y
\right)^{-s}\Gamma(-\nu - 1 + \mbox{$\frac{1}{2}$}s)\, .
\tselea{4.9}
\end{equation}
\noindent The integrand of (\tseref{4.9}) has simple poles in $s=-2n \;
(n=1,2,\ldots)$, $s=1$, $s=-2n +2\nu +2 \;(n= 0,1, \ldots, \nu)$
 and double pole in $s=0$. An important point in the former pole analysis
was the fact that $\zeta(s)$ has simple zeros in $-2m$ ($m>0$) and only
one simple pole in $s=1$. The former together with identity
\begin{displaymath}
\Gamma\left( \frac{x}{2} \right)\; \pi^{- \frac{x}{2}}\; \zeta(x)
= \Gamma \left( \frac{1-x}{2} \right) \; \pi^{ \frac{x-1}{2}} \;
\zeta(1-x)\, ,
\end{displaymath}
\noindent shows that no double pole except for $s=0$ is present in
(\tseref{4.9}). Now, we can close the contour to the left as the
value of the contour integral around the large arc is zero in the
limit of infinite radius (c.f.\tsecite{B} and\tsecite{GR}; formula
8.328.1). Using successively the Cauchy theorem we obtain
\begin{eqnarray}
&&\mbox{\hspace{-16mm}}\lefteqn{\frac{4\sqrt{\pi}\;\mbox{(\tseref{4.7})}}
{\Gamma(\mbox{$\frac{3}{2}$}+\nu)}}\nonumber \\
&&\mbox{\hspace{-11mm}}= \ \sum_{n=0}^{\nu} y^{2n - 2\nu -2} \;
\frac{\pi^{-2n +2\nu +2}(-n+\nu)!\; (-1)^{n} |B_{-2n + 2\nu
+2}|}{n! \;
(-2n +2\nu +2)!\; 2^{4n -4\nu -4}}\nonumber \\
&&\mbox{\hspace{-11mm}}+ \ \sum_{n=1}^{\infty} y^{2n} \;
\frac{\pi^{-2n} \;(2n)! \;\zeta(1+2n)\; (-1)^{n + \nu +1}}{n!\; (n
+ 1 + \nu)!
\;2^{4n-1}}\nonumber \\
&&\mbox{\hspace{-11mm}}+ \ y^{-1} \; \frac{\pi\; (-1)^{\nu + 1} \;
(\nu + 1)! \;2^{2\nu +3}}{(2\nu + 2)!} + \frac{2\; (-1)^{\nu +
1}}{(\nu + 1)!}\left\{ \mbox{ln} \left( \frac{y}{4\pi}\right) +
\gamma - \mbox{$\frac{1}{2}$} \sum_{k=1}^{\nu + 1}\frac{1}{k}
\right\}\, , \tselea{4.10}
\end{eqnarray}
\noindent where $B_{\alpha}$'s are the Bernoulli numbers. Let us mention
that for $\zeta(2n + 1)$ only numerical values are available.

\vspace{3mm}

Inserting (\tseref{4.10}) back to (\tseref{4.4}), we get for
${\cali{P}}(T) - {\cali{P}}(0)$
\begin{eqnarray}
&&{\cali{P}}(T) - {\cali{P}}(0) = (\mbox{\tseref{4.3}}) +
\frac{1}{3}\; I_{2}(m_{r}(T)) + \frac{\delta
m^{2}(T)}{4} \;I_{0}(m_{r}(T))\nonumber \\
&&\mbox{}\nonumber\\
&& \mbox{\hspace{5mm}}=\frac{T^{4}\; \pi^{2}}{90} -
\frac{T^{2}}{24} \left( m^{2}_{r}(T) - \frac{\delta m^{2}(T)}{2}
\right) + \frac{T\; m_{r}(T)}{4\; \pi} \left(
\frac{m^{2}_{r}(T)}{3} - \frac{\delta m^{2}(T)}{4} \right)\nonumber \\
&&\mbox{}\nonumber\\
&& \mbox{\hspace{5mm}} +\; \frac{m_{r}^{2}(T)\;m^{2}_{r}(0)}{32\;
\pi^{2}}\; \left(\mbox{ln}\left( \frac{m_{r}(0)}{T\; 4\pi}
\right) + \gamma -
\frac{1}{2} \right) - \frac{m^{4}_{r}(0)}{128 \; \pi^{2}} \nonumber \\
&&\mbox{}\nonumber\\
&& \mbox{\hspace{5mm}} - \; \sum_{n=1}^{\infty} \left(
m^{2}_{r}(T) - \mbox{$\frac{(n+2)}{2}$}\; \delta m^{2}(T)\right)
\; \frac{m_{r}^{2n+2}(T) \; \pi^{-2n-2}\; (2n)!\; \zeta (1+2n) \;
(-1)^{1+n}}{T^{2n}\;n!\;(n+2)!\;
2^{4n+4}}\, .\nonumber\\
\tselea{4.11}
\end{eqnarray}
%
%
\noindent Note that (\tseref{4.3}) cancelled against the same term
in $\frac{1}{3}\; I_{2}(m_{r}(T)) + \frac{\delta m^{2}(T)}{4}
\;I_{0}(m_{r}(T))$. One can see that (\tseref{4.11}) rapidly
converges for large $T$, so that only first four terms dominate at
sufficiently high temperature. The aforementioned terms come from
the poles nearby the straight line $(-i\infty + a, \; i\infty +a)$
(the more dominant contribution the closer pole). It is a typical
feature of the Mellin transform technique that integrals of type
\begin{displaymath}
\int_{0}^{\infty}dx \; g(x) \; \frac{1}{e^{xy}-1}\, ,
\end{displaymath}
\noindent can be expressed as an expansion which rapidly converges
for small $y$ (high--temperature expansion) or large $y$
(low--temperature expansion)\footnote{By the same token we get the
low--temperature expansion if the integral contour must be closed
to the right.}.

\vspace{3mm}

Expansion (\tseref{4.11}) is the sought result. To check its
consistency we will apply it to two important cases: high $T$ case
and $m_r(0) = 0$ case. Concerning the first case, note that for a
sufficiently large $T$ we can use the high--temperature expansion
of $\delta m^{2}(T)$ found in Appendix B. Inserting
(\tseref{bb52})  to (\tseref{4.11}) we obtain
\begin{eqnarray}
&&\mbox{\hspace{-8mm}}{\cali{P}}(T) - {\cali{P}}(0) =
\frac{T^{4}\; \pi^{2}}{90} - \frac{T^{2}\;m^{2}_{r}(T)}{24} +
\frac{T^{3}\;m_{r}(T)}{12 \pi}\nonumber
\\
&& \nonumber \\
&&\mbox{\hspace{-4mm}} + \ \frac{\lambda_{r}}{8}\left(
\frac{T^{4}}{144} - \frac{T^{3}\;m_{r}(T)}{24 \pi} +
\frac{T^{2}\;m^{2}_{r}(T)}{16 \pi^{2}}\right) + {\cali{O}}\left(
m^{4}_{r}(T)\; \mbox{ln}\left( \frac{m_{r}(T)}{T4 \pi }\right)
\right)\, . \tselea{pj1}
\end{eqnarray}
\vspace{1mm}
\noindent Up to a sign, the  result (\tseref{pj1}) coincides with
that found by Amelino--Camelia   and  Pi\tsecite{ACP}  for  the
effective potential\footnote{Let  us remind\tsecite{ACP,CJT,HS}
that from   the definition of $V_{eff}$ the thermodynamic pressure
is $-V_{eff}$. In order to obtain (\tseref{pj1}) from $V_{eff}$
in\tsecite{ACP}, one must subtract the zero temperature value  of
$V_{eff}$ and restrict oneself to   vanishing field expectation
value    and positive bare    mass squared.}. Actually,  they used
instead of  the $N \rightarrow \infty$ limit the  Hartree--Fock
approximation which is  supposed to  give the same $V_{eff}$ as
the leading $1/N$ approximation\tsecite{AC2}.

\vspace{3mm}

As for the second case, we may observe that our discussion of the
mass renormalization in Section 3.1 can be directly extended to
the case when $m_{r}(0) = 0$ (this does not apply to our
discussion of $\lambda_{r}$!). Latter can be also seen from the
fact that (\tseref{4.11}) is continuous in $m_{r}(0)=0$ (however
not analytic). The foregoing implies that the original massless
scalar particles acquire the thermal mass $m_{r}^{2}(T)= \delta
m^{2}(T) $ . From (\tseref{4.11}) one then may immediately deduce
the pressure for massless fields $\phi_{a}$ in terms of $\delta
m(T)$. The latter reads
\begin{eqnarray}
{\cali{P}}(T) - {\cali{P}}(0) &=& \frac{T^{4} \; \pi^{2}}{90}  -
\frac{T^{2}\; (\delta m(T))^{2}}{48} + \frac{T\; (\delta m(T))^{3}}{48
\; \pi} \nonumber \\
&& \mbox{} \nonumber \\
&+& \; \sum_{n=1}^{\infty} \frac{(\delta m(T))^{2n+4}
\;\pi^{-2n-2}\; (2n)!\; \zeta(1+2n)\; (-1)^{n+1}}{T^{2n}
\;(n-1)!\; (n+2)!\; 2^{4n+5}}\,
. \tselea{4.12}
\end{eqnarray}
\noindent This result is identical to that found by Drummond {\em
et al.} in\tsecite{ID1}.

\vspace{3mm}

A noteworthy observation is that when the energy of a thermal
motion is much higher then the mass of particles in the rest, then
the massive theory approaches the massless one. This is justified
in the first (high--temperature dominant) term of (\tseref{4.11})
and (\tseref{4.12}).  This term is nothing but a half of the black
body radiation pressure for photons\tsecite{GM,Cub} (photons have
two degrees of freedom connected with two transverse
polarizations). One could also obtain the temperature dominant
contributions directly from the Stefan--Boltzmann
law\tsecite{LW,GM,Cub} for the density energy (i.e. $\langle
\Theta^{00} \rangle$). The formal argument leading to this
statement is based on the noticing that at high energy
(temperature) the scalar field theory is approximately conformally
invariant, which in turn implies that the energy--momentum tensor
is traceless\tsecite{CCR}.  Taking into account the definition of
the hydrostatic pressure (\tseref{EMT24}), we can with a little
effort recover the leading high--temperature contributions for the
massive case.

\section{Conclusions} \label{C2}

In the present article we have clarified the status of hydrostatic
pressure in (equilibrium) thermal QFT. The former is explained in
terms of the thermal expectation value of the ``weighted"
space--like trace of the energy--momentum tensor $\Theta^{\mu
\nu}$. In classical field theory there is a clear microscopic
picture of the hydrostatic pressure which is further enhanced by a
mathematical connection (through the virial theorem) with the
thermodynamic pressure. In addition, it is the hydrostatic
pressure which can be naturally extended to a non--equilibrium
medium. Quantum theoretic treatment of the hydrostatic pressure is
however pretty delicate. In order to get a sensible, finite answer
we must give up the idea of total hydrostatic pressure. Instead,
thermal interaction pressure or/and interaction pressure must be
used (see (\tseref{ppp9}) and (\tseref{ppp10})). We have
established this result for a special case when the theory in
question is the scalar $\phi^{4}$ theory with $O(N)$ internal
symmetry; but it can be easily extended to more complex
situations. Moreover, due to a lucky interplay between the
conservation of $\Theta^{\mu \nu}$ and the space--time
translational invariance of an equilibrium (and $T=0$) expectation
value we can use the simple canonical (i.e. unrenormalized)
energy--momentum tensor. In the course of our treatment in Section
\ref{PE2} we heavily relied on the counterterm renormalization,
which seems to be the most natural when one discusses
renormalization of composite Green's functions. To be specific, we
have resorted to the minimal subtraction scheme which has proved
useful in several technical points.

\vspace{3mm}

We have   applied the prescriptions  obtained  for the QFT
hydrostatic pressure to $\phi^{4}$ theory in the--large $N$ limit.
The former has the undeniable advantage of being exactly soluble.
This is because of the fact  that  the large--$N$  limit
eliminates ``nasty"  classes  of diagrams in the thermal
Dyson--Schwinger expansion.  The survived  class  of diagrams
(superdaisy diagrams) can  be  exactly resumed, because the
(thermal)  proper self--energy  ${\vect{\Sigma}}$,  as  well as
the renormalized coupling   constant    $\lambda_{r}$    are
momentum independent. We have also stressed that the
$O(N)\;\phi^{4}$ theory in the large--$N$ limit is consistent only
if  we view it as an effective field theory. Fortunately, the
upper bound on the UV cut--off is truly huge,  and it does not
represent  any significant restriction. For the model at  hand the
resumed form of  the pressure with $m_{r}(0)=0$ was firstly
derived (in  the purely  thermodynamic pressure  context) by
Drummond {\em  et al.} in\tsecite{ID1}.  We  have  checked, using
the prescription (\tseref{ppp9}) for   the thermal interaction
pressure, that their  results are in agreement with  ours. The
former is  a nice vindication of  the validity of  the  virial
theorem for the QFT system at hand. In this connection we should
perhaps mention that the latter is by no means obvious. For
example, for quantized gauge fields the conformal (trace) anomaly
may even invalidate the virial theorem\tsecite{LW}. The fact that
this point is indeed non--trivial is illustrated on the QCD case
in\tsecite{DHLR4}.
%

\vspace{3mm}

The expression for the pressure obtained was in a suitable form
which allowed us to take advantage of the Mellin transform
technique. We were then able to write down the high--temperature
expansion for the pressure in $D=4$ (both for massive and massless
fields) in terms of renormalized masses $m_{r}(T)$ and $m_{r}(0)$.
We have explicitly checked that all UV divergences present in the
individual thermal diagrams ``miraculously" cancel in accordance
with our analysis of the composite operators in Section \ref{PE2}.

\section*{Acknowledgements}

I am indebted to P.V.~Landshoff for reading the manuscript and for
invaluable discussion. I am also grateful to N.P.~Landsman,
H.~Osborn, R.~Jackiw and H.J.~Schnitzer for useful discussions.
Finally I would like to thank the Fitzwilliam College of Cambridge
and the Japanese Society for Promotion of Science for financial
supports.

\appendix
\section{Appendix}

In this Appendix we give some details of the derivation of
Eq.(\tseref{b10}). We particularly show that the surface integrals
arisen during the transition from (\tseref{b9}) to (\tseref{b10})
mutually cancel among themselves. As usual, the integrals will be
evaluated for integer values of $D$ and corresponding results then
analytically continued to a desired (generally complex) $D$.

\vspace{3mm}

The key quantity in question is
\begin{eqnarray}
&+& \frac{1}{2}\; \int \frac{d^{D}q}{(2\pi)^{D-1}} \left(
\frac{2{\vect{q}}^{2}}{(D-1)}
\right) \; \frac{\varepsilon(q_{0})}{e^{\beta
q_{0}}-1}\;\delta(q^{2}-m^{2}_{r}(T))\nonumber \\
&-& \frac{1}{2}\; \int \frac{d^{D}q}{(2\pi)^{D-1}} \left(
\frac{2{\vect{q}}^{2}}{(D-1)} \right)\;
\delta^{+}(q^{2}-m^{2}_{r}(0))\, . \tselea{a1}
\end{eqnarray}
\noindent Applying Green's theorem (i.e. integrating by parts with
respect to ${\vect{q}}$) on (\tseref{a1}) one finds
\begin{eqnarray}
&&\mbox{\hspace{-6mm}}\mbox{(\tseref{a1})}\ = \
{\cali{N}}_{T}(m^{2}_{r}(T)) -
{\cali{N}}(m^{2}_{r}(0))\nonumber\\
 &&\mbox{\hspace{-3mm}}+ \ \lim_{R \rightarrow \infty} \;
\frac{1}{2(D-1)} \int \frac{dq_{0}}{(2\pi)^{D-1}}\; \int_{\partial
S^{D-2}_{R}}
d{\vect{s}}\;{\vect{q}}\; \theta(q^{2}-m^{2}_{r}(T))\; \theta(q_{0})\;
\left( \frac{2}{e^{\beta q_{0}}-1}+1 \right)\nonumber \\
&&\mbox{\hspace{-3mm}}- \ \lim_{R \rightarrow \infty} \;
\frac{1}{2(D-1)} \int \frac{dq_{0}}{(2\pi)^{D-1}}\; \int_{\partial
S^{D-2}_{R}} d{\vect{s}}\;{\vect{q}}\;
\theta(q^{2}-m^{2}_{r}(0))\; \theta(q_{0})\, . \\
&~& \nonumber \tseleq{a2}
\end{eqnarray}
\noindent As usual, ${\vect{a}}{\vect{b}} =
\sum_{i=1}^{D-1}{\vect{a}}_{i}{\vect{b}}_{i}$ and $S^{D-2}_{R}$ is
a $(D-2)$--sphere with the radius $R$. The expressions for
${\cali{N}}_{T}$ and ${\cali{N}}$ are done by (\tseref{b101}).

\vspace{3mm}

With the relation (\tseref{a2}) we can show that the surface terms
cancel in the large $R$ limit. Let us first observe that
\begin{eqnarray}
&&\lim_{R \rightarrow \infty} \; \int
\frac{dq_{0}}{(2\pi)^{D-1}}\; \int_{\partial S^{D-2}_{R}}
d{\vect{s}}\;{\vect{q}}\; \theta(q^{2}-m^{2}_{r}(T))\;
\frac{2 \theta(q_{0})}{e^{\beta q_{0}}-1}\nonumber \\
&=& \lim_{R \rightarrow \infty}\;
\frac{2\pi^{\frac{D-1}{2}}R^{D-1}}{\Gamma \left(\frac{D-1}{2}\right)} \;
\int \frac{dq_{0}}{(2\pi)^{D-1}} \; \theta(q_{0}^{2}-R^{2}-m^{2}_{r}(T))\;
\frac{2 \theta(q_{0})}{e^{\beta q_{0}}-1}\nonumber \\
&=& \lim_{R \rightarrow \infty}
\frac{\pi^{\frac{1-D}{2}}R^{D-1}}{2^{D-2}\Gamma \left(
\frac{D-1}{2} \right)}\int_{\sqrt{R^{2}+m_{r}^{2}(T)}}^{\infty}
dq_{0}\; \frac{2}{e^{\beta q_{0}}-1} \; = \;0\, . \tseleq{a3}
\end{eqnarray}
\noindent In 2--nd line we have exploited Gauss's theorem and in
the last line we have used L'H{\^{o}}pital's rule as the
expression is in the indeterminate form $0/0$. The remaining
surface terms in (\tseref{a2}) read
\begin{eqnarray}
&& \lim_{R \rightarrow \infty} \int
\frac{dq_{0}}{(2\pi)^{D-1}} \; \int_{\partial S^{D-2}_{R}}
d{\vect{s}}\;{\vect{q}}\; \left\{ \theta(q^{2}-m^{2}_{r}(T)) -
\theta(q^{2}-m^{2}_{r}(0)) \right\}\; \theta(q_{0}) \nonumber\\
&=& \lim_{R \rightarrow \infty}\;
\frac{\pi^{\frac{1-D}{2}}R^{D-1}}{2^{D-2}\Gamma \left(
\frac{D-1}{2} \right)}\; \left\{
\int_{\sqrt{R^{2}+m_{r}^{2}(T)}}^{\infty} -
\int_{\sqrt{R^{2}+m_{r}^{2}(0)}}^{\infty} \right\}dq_{0}\; = \;
0\, . \tseleq{a4}
\end{eqnarray}
\noindent The last identity follows either by applying
L'H{\^{o}}spital's rule or by a simple transformation of variables
which renders both integrals inside of $\{ \ldots \}$ equal.
Expressions on the last lines in (\tseref{a3}) and (\tseref{a4})
can be clearly (single--valuedly) continued to the region
$\mbox{Re}D > 1$ as they are analytic there. We thus end up with
the statement that
\begin{displaymath}
\mbox{(\tseref{a1})}= {\cali{N}}_{T}(m^{2}_{r}(T))-
{\cali{N}}(m^{2}_{r}(0))\, .
\end{displaymath}
%

\section{Appendix}

In this appendix we shall derive the high--temperature expansion
of the mass shift $\delta m^{2}(T)$ in the case when fields
$\phi_{a}$ are massive (i.e. $m^{2}_{r}(0) \not= 0$).

\vspace{3mm}

Consider Eqs.(\tseref{c1}) and (\tseref{m12}). If we combine them
together, we get easily the following transcendental equation for
$\delta m^{2}(T)$
\begin{equation}
\delta m^{2}(T) = \lambda_{0} \left\{ {\cali{M}}(m^{2}_{r}(T)) -
{\cali{M}}(m^{2}_{r}(0)) + \frac{1}{2} I_{0}(m^{2}_{r}(0) + \delta
m^{2}(T))\right\}\, . \tseleq{bb1}
\end{equation}
\noindent Here ${\cali{M}}$ and $I_{0}$ are done by (\tseref{b45})
and (\tseref{c14}), respectively.

\vspace{3mm}

Now, both $\lambda_{0}$ and ${\cali{M}}$ are divergent in $D=4$.
If we reexpress $\lambda_{0}$ in terms of $\lambda_{r}$,
divergences must cancel, as $\delta m^{2}(T)$ is finite in $D=4$.
The latter can be easily seen if we Taylor expand ${\cali{M}}$,
i.e.
\begin{equation}
{\cali{M}}(m^{2}_{r}(T)) = {\cali{M}}(m^{2}_{r}(0)) + \delta
m^{2}(T) \; {\cali{M}}'(m^{2}_{r}(0)) +
\hat{{\cali{M}}}(m^{2}_{r}(0); \delta m^{2}(T))\, . \tseleq{bb2}
\end{equation}
\noindent Obviously, $\hat{{\cali{M}}}$ is finite in $D=4$ as
${\cali{M}}$ is quadratically divergent. Inserting (\tseref{bb2})
to (\tseref{bb1}) and employing Eq.(\tseref{m146}) we get
\begin{equation}
\delta m^{2}(T) = \lambda_{r} \left\{
\hat{{\cali{M}}}(m^{2}_{r}(0); \delta m^{2}(T)) + \frac{1}{2}
I_{0}(m^{2}_{r}(0) + \delta m^{2}(T))\right\}\, . \tseleq{bb3}
\end{equation}
\noindent This is sometimes referred to as the renormalized gap
equation. In order to determine $\hat{{\cali{M}}}$ we must go back
to (\tseref{bb2}). From the former we read off that
\begin{eqnarray}
&&\hat{{\cali{M}}}(m^{2}_{r}(T); \delta m^{2}(T))\nonumber \\
&&\nonumber \\
&& \mbox{\hspace{0.8cm}}= {\cali{M}}(m^{2}_{r}(T)) -
{\cali{M}}(m^{2}_{r}(0)) - \delta m^{2}(T)\;
{\cali{M}}'(m^{2}_{r}(0))\nonumber \\
&& \mbox{\hspace{0.8cm}}= \frac{\Gamma(1-
\frac{D}{2})}{2(4\pi)^{\frac{D}{2}}} \left\{
(m^{2}_{r}(T))^{\frac{D}{2}-1} -
(m^{2}_{r}(0))^{\mbox{$\frac{D}{2}$}-1} - \delta
m^{2}(T)(\frac{D}{2}-1)\; (m^{2}_{r}(0))^{\frac{D}{2}-2} \right\}
\nonumber \\
&& \mbox{\hspace{0.8cm}}
 \stackrel{D \rightarrow 4}{=}  \frac{1}{32
\pi^{2}} \left\{ m^{2}_{r}(T) \; \mbox{ln}\left(
\frac{m^{2}_{r}(T)}{m^{2}_{r}(0)}\right) - \delta m^{2}(T)
\right\}\, . \tseleq{bb41}
\end{eqnarray}
\noindent So
\begin{equation}
\delta m^{2}(T) = \lambda_{r} \left\{ \frac{(m^{2}(0) + \delta
m^{2}(T))\;\mbox{ln}\left( 1 + \frac{\delta
m^{2}(T)}{m^{2}_{r}(0)}\right) - \delta m^{2}(T)}{32 \pi^{2}} +
\frac{1}{2}I_{0} \right\}\, . \tseleq{bb4}
\end{equation}
\noindent Analogous relation was also derived in\tsecite{FS1}
where authors used finite temperature renormalization group. In
the latter the zero--momentum  renormalization prescription was
utilized. Eq.(\tseref{bb4}) was firstly obtained and numerically
solved in\tsecite{ID1}. It was shown that the solution is double
valued. The former behavior was also observed in the effective
action approach. Namely by Abbott {\em{et al.}}\tsecite{AKS} at
$T=0$, and by Bardeen and Moshe\tsecite{BM} at both $T=0$ and
$T\not=0$. The relevant solution is only that which fulfils the
consistency condition $\delta m^{2}(T) \rightarrow 0$ when $T
\rightarrow 0$. For such a solution it can be shown
(c.f.\tsecite{ID1}, Fig.3) that $\frac{\delta
m^{2}(T)}{m^{2}_{r}(0)}\ll 1$ for a sufficiently high $T$. So the
high--temperature expansion of (\tseref{bb4}) reads
\begin{eqnarray}
\delta m^{2}(T) &=& \lambda_{r}\left\{ \frac{\frac{(\delta
m^{2}(T))^{2}}{2m^{2}_{r}(0)} - \frac{(\delta m^{2}(T))^{3}}{6
m^{4}_{r}(0)} + \frac{(\delta m^{2}(T))^{4}}{12m^{6}_{r}(0)} +
\ldots
}{32 \pi^{2}}  + \frac{1}{2}I_{0} \right\}\nonumber \\
&&\nonumber \\
&\doteq& \frac{\lambda_{r}}{2} I_{0} = \frac{\lambda_{r}T^{2}}{24}
- \frac{\lambda_{r}m_{r}(T)}{8 \pi}\;T + {\cali{O}}\left(
m_{r}^{2}(T) \;\mbox{ln}\left(\frac{m_{r}(T)}{T4 \pi} \right)
\right. \, . \tselea{bb52}
\end{eqnarray}

\end{document}